\documentclass[prb,aps,reprint,superscriptaddress,titlepage,floatfix,showkeys]{revtex4-2}
\usepackage{graphicx}
\usepackage{nicefrac}
\usepackage{amsfonts}
\usepackage[version=3]{mhchem}
\usepackage{amssymb}
\usepackage{amsmath} 
\usepackage{subfigure}
\usepackage{multirow} 
\usepackage{tabularx} 
\usepackage{array}
\usepackage{units}
\usepackage{braket}
\usepackage{bm,times}
\usepackage{booktabs}
\usepackage{enumitem}
\usepackage{mathtools}
\usepackage{epstopdf}
\usepackage[dvipsnames]{xcolor} 
\usepackage{threeparttable}
\usepackage[colorlinks = true, linkcolor = blue, urlcolor  = blue, citecolor = blue, anchorcolor = blue]{hyperref}

\renewcommand{\Im}{\operatorname{{\mathrm Im}}}

\def\be{\begin{equation}}
\def\ee{\end{equation}}
\def\bea{\begin{eqnarray}}
\def\eea{\end{eqnarray}}

\def\deg{$^\circ$}

\def\asi{{\it a}-Si}
\def\age{{\it a}-Ge}
\def\csi{{\it c}-Si}
\def\dcsi{{\it dc}-Si}

\def\Ihat{\hat{I}}
\def\Hhat{\hat{H}}
\def\Ghat{\hat{G}}

\begin{document}
\title{
Extended-range order in tetrahedral amorphous semiconductors: The case of 
amorphous silicon}

\author{Devilal Dahal}
\email{devilal.dahal@usm.edu}
\affiliation{Department of Physics and Astronomy, The University of
Southern Mississippi, Hattiesburg, Mississippi 39406, USA}

\author{Stephen R. Elliott}
\email{stephen.elliott@chem.ox.ac.uk}
\affiliation{Physical and Theoretical Chemistry Laboratory,
University of Oxford, Oxford OX1 3QZ, United Kingdom}

\affiliation{Department of Chemistry, University of Cambridge,
Cambridge CB2 1EW, United Kingdom}

\author{Parthapratim Biswas}
\email[Corresponding author: ]{partha.biswas@usm.edu}
\affiliation{Department of Physics and Astronomy, The University of
Southern Mississippi, Hattiesburg, Mississippi 39406, USA}

\begin{abstract}
This paper reports the presence of extended-range ordering
in the atomic pair-correlation function of amorphous 
silicon ({\it a}-Si) using ultra-large atomistic models 
obtained from Monte Carlo and molecular-dynamics 
simulations. The extended-range order manifests itself 
in the form of radial oscillations, on the length scale 
of 20--40 {\AA}, which are examined by directly analyzing 
the radial distribution of atoms in distant coordination shells 
and comparing the same with those from a class of 
partially-ordered networks of Si atoms and disordered 
configurations of crystalline silicon from an 
information-theoretic point of view. 
The study suggests that the extended-range radial oscillations 
principally originate from the propagation of radial 
ordering from the first few atomic shells to a distance 
of up to 40 {\AA}. The effect of 
these oscillations on the first sharp diffraction 
peak (FSDP) in the structure factor is addressed 
by obtaining a semi-analytical expression for the static 
structure factor of {\it a}-Si, and calculating an estimate of the error 
of the intensity of the FSDP associated with the 
truncation of radial information from distant shells. 
The results indicate that the extended-range
oscillations do not have any noticeable effects on
the position and intensity of the FSDP, which are 
primarily determined by the medium-range atomic 
correlations of up to a length of 20 {\AA} in 
amorphous silicon.
\end{abstract}  

\maketitle

\section{Introduction}
The structure of amorphous silicon ({\asi}) is well 
represented by the continuous random network (CRN) 
model of Zachariasen~\cite{Zach1932}. The CRN model 
of {\asi} suggests that each atom is bonded to 
four neighboring Si atoms, which form an approximate 
tetrahedral atomic arrangement in the amorphous environment. 
The network is topologically distinct from its crystalline 
counterpart ({\csi}) owing to the presence of 5-member 
and 7-member rings. In addition, a considerable number 
of hexagonal rings and a few higher-member rings are also 
present in the amorphous network. The 
pair-correlation function (PCF) of {\asi} obtained from 
CRN models indicates that radial correlations 
typically extend up to a distance of 15 {\AA}. Although 
the actual structure of laboratory-grown samples of 
{\asi} may differ from this simple CRN picture, except 
for a few properties, the CRN model provides an overall 
good description of structural, electronic, and 
vibrational properties of {\asi} that mostly rely on 
the short-range order ($\approx$ 5 {\AA}) and, to a 
lesser extent, the medium-range order ($\approx$ 5--20 
{\AA}) of the network. 

Although the structure of {\asi} has been extensively studied 
by using computer-generated models on the radial 
length scale of 10--15 {\AA}, there exist only 
a few studies~\cite{Elliott:1991, Biswas:2020, Biswas:2015, Dahal:2019} that 
discuss the network structure of {\asi} on the 
medium-range length scale of 20 {\AA} and 
beyond. This is partly due to the fact that structural 
and electronic properties of {\asi} are generally found to be not 
particularly dependent on the medium-range structure 
beyond 15 {\AA} and in part to the computational 
complexity of conducting quantum-mechanical calculations, 
using density-functional theory (DFT), for large 
models.  However, this observation does not necessarily 
imply that no medium-range structure exists in 
{\asi}~\cite{Gibson:1997}.  In this paper, we address 
this aspect of the problem by studying the network structure 
of {\asi} using atomistic models of sizes 21,952 and 
400,000 atoms. In particular, we examine two important 
aspects of the medium- and extended-range structures 
of {\asi} that have been sparsely reported in the literature. 
The first problem involves the presence of weak but 
noticeable radial oscillations in the PCF at 
distances of 20--40 {\AA}. 
This was first reported by Uhlherr and Elliott~\cite{Uhlherr:1994} 
and it was given the name extended-range oscillations 
in the PCF of {\asi}. The second issue is directly related 
to the first and it concerns the effect of the medium-range 
order beyond 15 {\AA}, and possibly the extended-range 
order, on the first sharp diffraction peak (FSDP) 
of {\asi}. The latter corresponds to the first peak of 
the static structure factor~\cite{Xie:2013}, 
$S(Q)$, at $Q$ = 1.99 {\AA}$^{-1}$ in {\asi}. In the following, we 
use the term medium-range order (MRO) to imply 
ordering on the length scale of 5--20 {\AA}, whereas the 
term ERO indicates structural ordering beyond 20 {\AA}, 
including extended-range oscillations. 

The role of the medium-range order (MRO) in amorphous networks 
has been studied extensively in an effort to understand 
structure-property relationships in network-forming 
glasses, for example, 
oxides~\cite{Aniya:2004,Shyam:2016,Mei:2008,Salmon:2006,Sampath:2003} 
and chalcogenides~\cite{Vashishta:1989, Phillips:1981, 
Lucovsky:1987,Iyetomi:1991,Armand:1992}. The MRO in these 
systems typically manifests itself as the FSDP, and the position, 
width, and intensity of the FSDP characterize the 
length scale associated with the MRO. 
The results from numerous experimental~\cite{Sheng:2006, Yang:2021, Ma:2009, Cormier:1998, Salmon:2007, Mei:2008, Hazra:2004} 
and computational studies~\cite{Elliott:1991, Hirata:2011, Sharma:2006, Tanaka:1998, Nishio:2013, Du:2006, Deringer:2021}
indicate that the MRO/ERO in glassy systems can extend up 
to a distance of 30 {\AA} and that it can play an 
important role in determining a number of materials 
properties of network-forming glasses. By contrast, results 
for tetrahedrally-bonded elemental amorphous semiconductors, 
such as {\asi} and {\age}, are 
few and far between. Uhlherr and Elliott~\cite{Uhlherr:1994} 
studied the presence of extended-range oscillations in 
{\asi} by analyzing experimental neutron-diffraction data 
of Fortner and Lannin~\cite{Fortner:1989} and the 
pair-correlation data obtained from atomistic models 
of size 13,824 atoms.~\cite{Holender:1991} The authors 
concluded, via the Fourier inversion of the structure 
factor in the vicinity of the FSDP region, that the 
radial oscillations can extend to at least 35 {\AA} and 
that it arises from the propagation of second-neighbor 
radial atomic correlations. Recently, Roorda et al.~\cite{Roorda:2012} 
reported the presence of ERO in amorphous Si/Ge 
using x-ray diffraction measurements at high 
resolution. The PCF obtained in their study from 
the Fourier transform of diffraction data shows the 
presence of ERO beyond 20 {\AA} in both {\asi} 
and {\age} samples.  The authors also noted 
that the (spatial) periodicity and decay length of 
the MRO/ERO increase upon thermal annealing. 
In view of these observations, the main task of the 
present study is to examine the presence of the ERO 
in large realistic models of {\asi} by a direct analysis of the 
pair-correlation function and their partial 
counterparts associated with distant coordination 
shells of amorphous silicon. 

The rest of the paper is arranged as follows. In 
section II, we have provided a description of 
the computational methods employed here to generate 
atomistic models of {\asi} and a set of partially-ordered 
networks of Si atoms. This is followed by results 
and discussion in section III. The origin of the 
ERO is addressed from a real-space point of view 
of the network structure of amorphous silicon. 
The relation between the ERO and structure of the 
FSDP is also examined in this section by constructing 
a semi-empirical expression for the structure factor 
of {\asi} in the Gaussian approximation. This is 
followed by conclusions of our work in section IV.

\section{Computational method}
The present study involves the use of three different 
sets of models. The first set consists of {\asi} models 
obtained from using the Wooten-Winer-Weaire (WWW)~\cite{Wooten:1985,Barkema:2000} 
algorithm. The second set comprises {\asi} 
models produced from large-scale molecular-dynamics 
(MD) simulations. 
The third set includes three different types 
of partially-ordered networks of Si atoms, denoted by 
M1, M2, and M3. These are not realistic models of 
{\asi} but have varying degrees of radial 
ordering in the respective PCF up to a radial length 
of 6 {\AA}. In order for the ERO 
to manifest in the PCF of {\asi} at radial distances 
of 20--40 {\AA}, it is necessary for the models to be 
sufficiently large, consisting of a few tens to several 
tens of thousands of atoms. To this end, the sizes 
of the models were chosen to be 21,952 atoms and 400,000 
atoms, which suffice to establish an unambiguous 
presence of the ERO in the PCF.  In this study, we 
generated a set of three independent M1/M2/M3 and 
WWW models and three MD models for the purpose of 
configurational averaging of data. 

The MD models were produced by initially placing 400,000 Si atoms 
randomly in a cubic simulation cell of length 202.4 {\AA}, 
so that no two atoms could be at a distance of less 
than 2 {\AA}.  The mass density of the models corresponds to 
2.26 g.cm$^{-3}$, which is close to the 
experimental value~\cite{Custer:1994, Laaziri:1999} of the 
{\asi} density, 2.25--2.28 g.cm$^{-3}$, depending upon the method 
of preparation and experimental conditions. The modified 
Stillinger-Weber potential~\cite{Vink:2001,SW:1985} was 
used to calculate the total energy and forces and the 
velocity-Verlet algorithm was employed to integrate the 
equations of motion in canonical ensembles. The initial 
temperature was set at 1800 K and the system was equilibrated 
for 20 ps at 1800 K. The temperature was then gradually decreased, by using 
a chain of Nos\'{e}-Hoover thermostats~\cite{Nose:1984, Hoover:1985},
from 1800 K to 300 K at an average cooling rate of 
$5\times10^{12}$ K/s.  The final 
structures from the MD simulations were further subjected 
to geometry optimization using the limited-memory BFGS 
algorithm, as described by~\citet{Ray:2018}.  Atomic 
configurations 
were collected during the course of simulations once 
the configurations satisfied a set of convergence properties, 
involving a minimum value of the width of the bond-angle 
distribution and the number of 4-fold-coordinated 
atoms in the network.  

The second set of models were produced by using the 
WWW method. Here, we employed the modified version 
of the algorithm, developed by Barkema and 
Mousseau~\cite{Barkema:2000}. The method 
essentially consists of the following steps: 1) 
Generate a random configuration and construct a 
neighbor list of atoms using an appropriate cutoff 
value, such that the network is tetravalent as far 
as the list is concerned; 2) Employ the WWW bond-switching 
algorithm~\cite{Wooten:1985, Barkema:2000} to produce 
a new configuration and accept or reject the configuration 
upon local relaxation of the network via the Monte 
Carlo method. The bond-switching procedure largely 
maintains the tetravalent character of the atomic 
network during simulations, and local relaxations 
were performed by using the nearest-neighbor-based Keating 
potential~\cite{Keating:1966}; 3) Relax the resulting 
configuration from step 2 at a regular but infrequent 
interval to include the structural information from 
beyond the first shell of neighbors, by using a 
generalization of Weber's adiabatic bond-charge 
model~\cite{Weber:1977}. For a description 
of the method, see \citet{Barkema:2000}. 

In addition to the WWW and MD models of {\asi}, we have 
also generated a set of disordered networks, M1--M3, 
of Si atoms. These networks are partially ordered and they can be 
classified by the degree of radial correlations present 
in the respective PCF. Specifically, M1 models are 
highly disordered and have very little or no radial 
correlations in the PCF. By contrast, M2 models are 
characterized by the presence of a well-defined 
first peak and radial correlations up to 3 {\AA}. 
Likewise, M3 models exhibit radial correlations 
up to 6 {\AA} with a pristine first peak and a 
part of the second peak, with a well-defined gap 
between the peaks. The M2 and M3 models were 
generated by adding one atom at a time in the 
simulation cell so that the addition of each atom 
satisfied a set of geometric constraints in order 
to produce radial correlations up to a length 
of 4 {\AA} and 6 {\AA}, respectively. The sizes 
of the WWW and M1/M2/M3 models were chosen to be 
21,952 atoms, with a cubic supercell of linear size 
77.03 {\AA}.

Apart from the WWW, MD, and M1 to M3 models, we have 
also employed a number of disordered amorphous 
silicon ({\it{da}}-Si) and disordered crystalline silicon ({\it{dc}}-Si) 
configurations in this study. These configurations 
were produced by including structural disorder in 
pristine {\asi} and diamond {\csi} structures 
via random displacements of atoms, using $r_{i, \alpha} \to 
r_{i, \alpha} + \sigma\,p_{i,\alpha}$, from their 
original positions. Here, $r_{i,\alpha}$ is the 
$\alpha$-th component ($\alpha = x/y/z$) of the 
atomic position at site $i$, $\sigma$ is the maximum 
value of the atomic displacement in {\AA}, and 
$p_{i,\alpha}$ is a random number, which is uniformly 
distributed between -1 and +1. The values of $\sigma$ 
were chosen from 0.2 {\AA} to 1.2 {\AA}, 
which correspond to a distortion of the Si--Si bond 
length by 8--51\% from its average/ideal 
value of 2.36 {\AA} in {\asi}/{\csi}. It may be noted that 
a value of $\sigma$ of the order of 0.3 {\AA} 
satisfies the Lindemann's criterion of melting, 
producing liquid-like structures of {\asi} and 
{\csi}. Thus, the {\dcsi} configurations with $\sigma 
\gg$ 0.3 {\AA} are considerably disordered 
compared to their counterparts with $\sigma 
\le $ 0.3 {\AA}.   

Given a distribution of atoms in a disordered 
network, the structure factor can be obtained 
from the Fourier transform of the reduced 
PCF, $G(r)$. Assuming that the distribution of atoms 
in the network is homogeneous and isotropic, 
the structure factor, $S(Q)$, is given by,
\bea
S(Q) & = & 1 + \frac{4\pi n_0}{Q} \int_0^{\infty} r[g(r)-1] \sin(Qr)\, dr \notag \\
     & \approx & 1 + \frac{1}{Q} \int_0^{R_c}  G(r) \sin(Qr)\, dr, 
\label{eq1}
\eea
where $g(r)$ is the conventional pair-correlation function 
(PCF), $G(r)= 4\pi n_0 \, r\,[g(r)-1]$ is known as the 
reduced PCF, and $n_{0}$ is the average number density 
of the system. For finite-size models, the upper limit 
of the integral can be replaced by $R_c = L/2$ by 
using the periodic boundary conditions, provided 
$g(r) \rightarrow 1$ as $r \to R_c$. We shall see later that 
this condition is amply satisfied by models for which 
$R_c$ is of the order of 20 {\AA}. 

\section{Results and Discussion} 
\subsection{Extended-range oscillations in the PCF of {\asi} }
\begin{figure}[t!]
\centering
\includegraphics[width=0.45\textwidth]{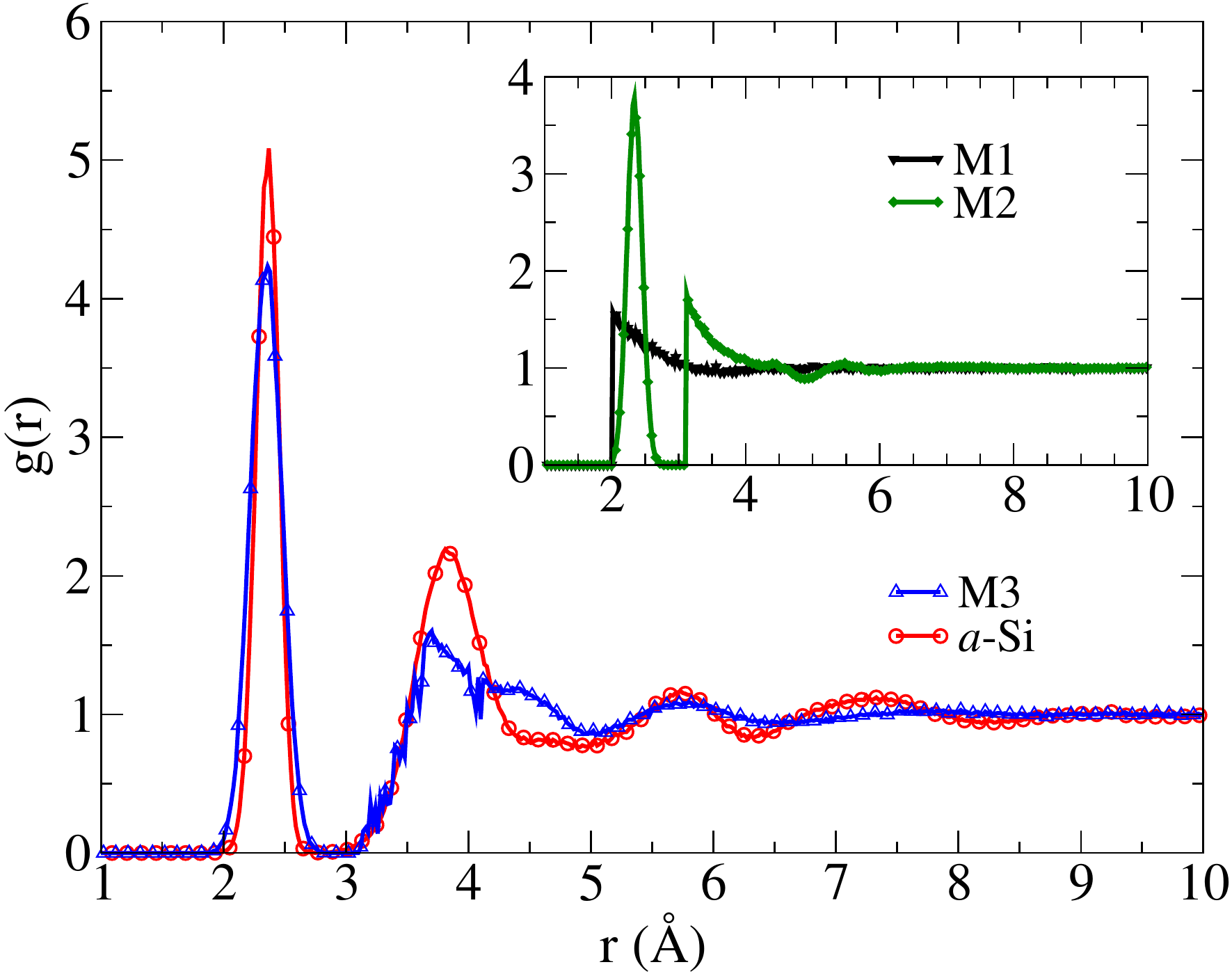}
\caption{
The pair-correlation functions of three partially-ordered 
models (M1 to M3) of Si atoms, showing radial correlations 
up to a length of 6 {\AA}. The results for {\asi} (WWW models) 
are shown for comparison with that for the M3 model. The 
size of the models corresponds to 21,952 atoms and the PCF 
data were averaged over three independent configurations 
for each model. 
}
\label{rdf_M}
\end{figure}

We begin by establishing the unambiguous presence of 
radial oscillations in the PCF of {\asi} at a distance 
of 20--40 {\AA}. Since the calculation of the PCF 
beyond 20 {\AA} requires sufficiently large models 
of {\asi}, we first examine the large MD models, 
consisting of 400,000 atoms. Thereafter, we proceed to 
determine the origin of these oscillations by 
analyzing the three-dimensional network 
structure of these 400,000-atom models and a set of 
21,952-atom models obtained from the WWW method. 
The results from these models will be compared with 
the same from the partially-ordered networks, M1 to 
M3, having varying degrees of radial ordering up 
to a distance of 6 {\AA}. 
The PCFs of the partially-ordered networks, from M1 
to M3, are shown in Fig.\,\ref{rdf_M}, along with 
the results from the 21,952-atom WWW models of 
{\asi}. It is evident from the plots that the M2 
and M3 models show radial correlations of 
up to 4 {\AA} and 6 {\AA}, respectively. The 
M1 models, on the other hand, exhibit small 
radial correlations up to 3 {\AA}, which mostly 
originate from the imposed constraint of a minimum 
separation distance of 2 {\AA} between any two 
atoms in the network. 

\begin{figure}[t!]
\centering
\includegraphics[width=0.45\textwidth]{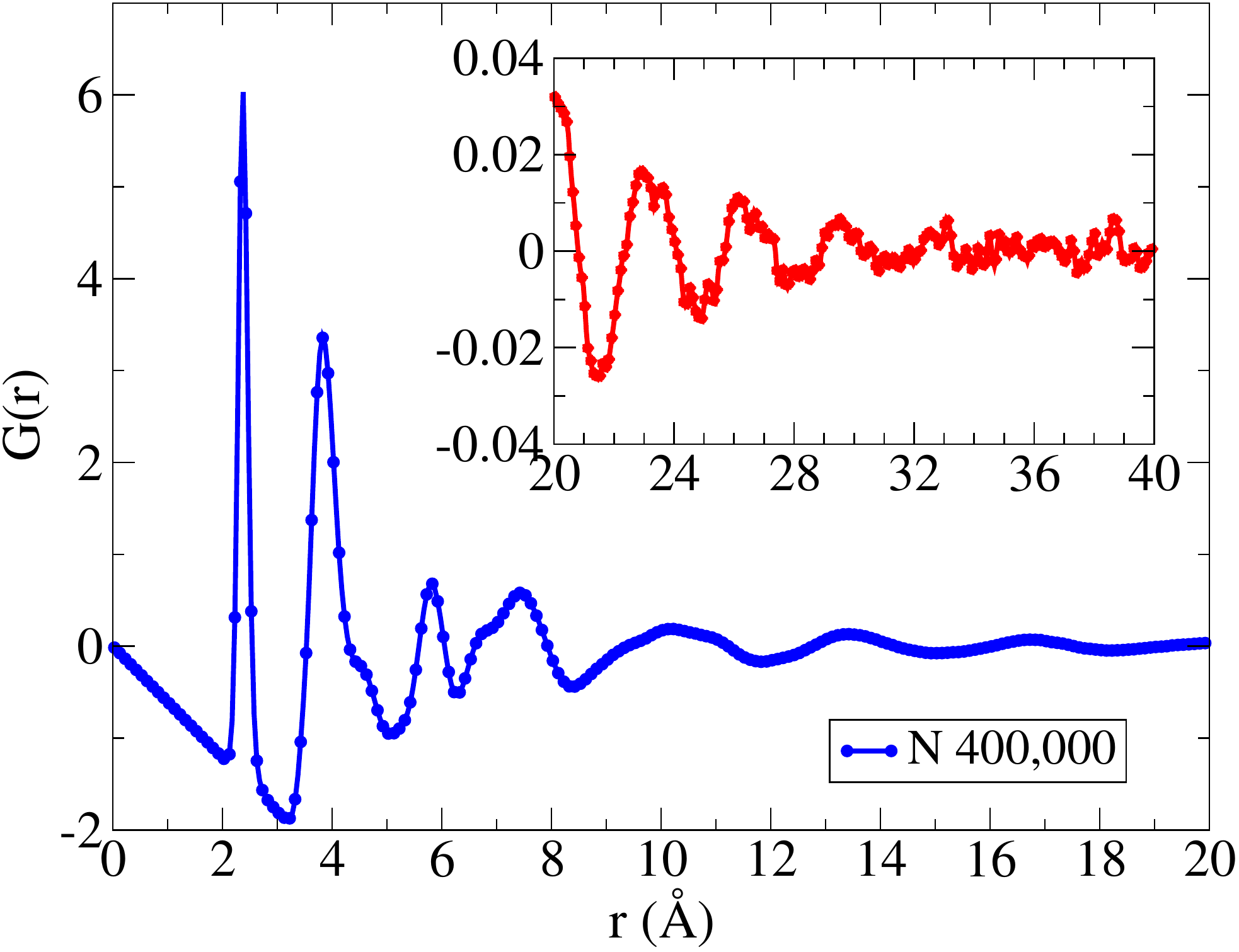} 
\caption{
The reduced pair-correlation function, $G(r)$, of {\asi}, 
obtained from a configurational averaging of three large 
MD models of size 400,000 atoms. The inset shows 
the presence of radial oscillations up to 40 {\AA}, 
which are known as the extended-range oscillations 
in {\asi}. 
}
\label{rdf_SI}
\end{figure}

\begin{figure}[t!]
\centering
\includegraphics[width=0.45\textwidth]{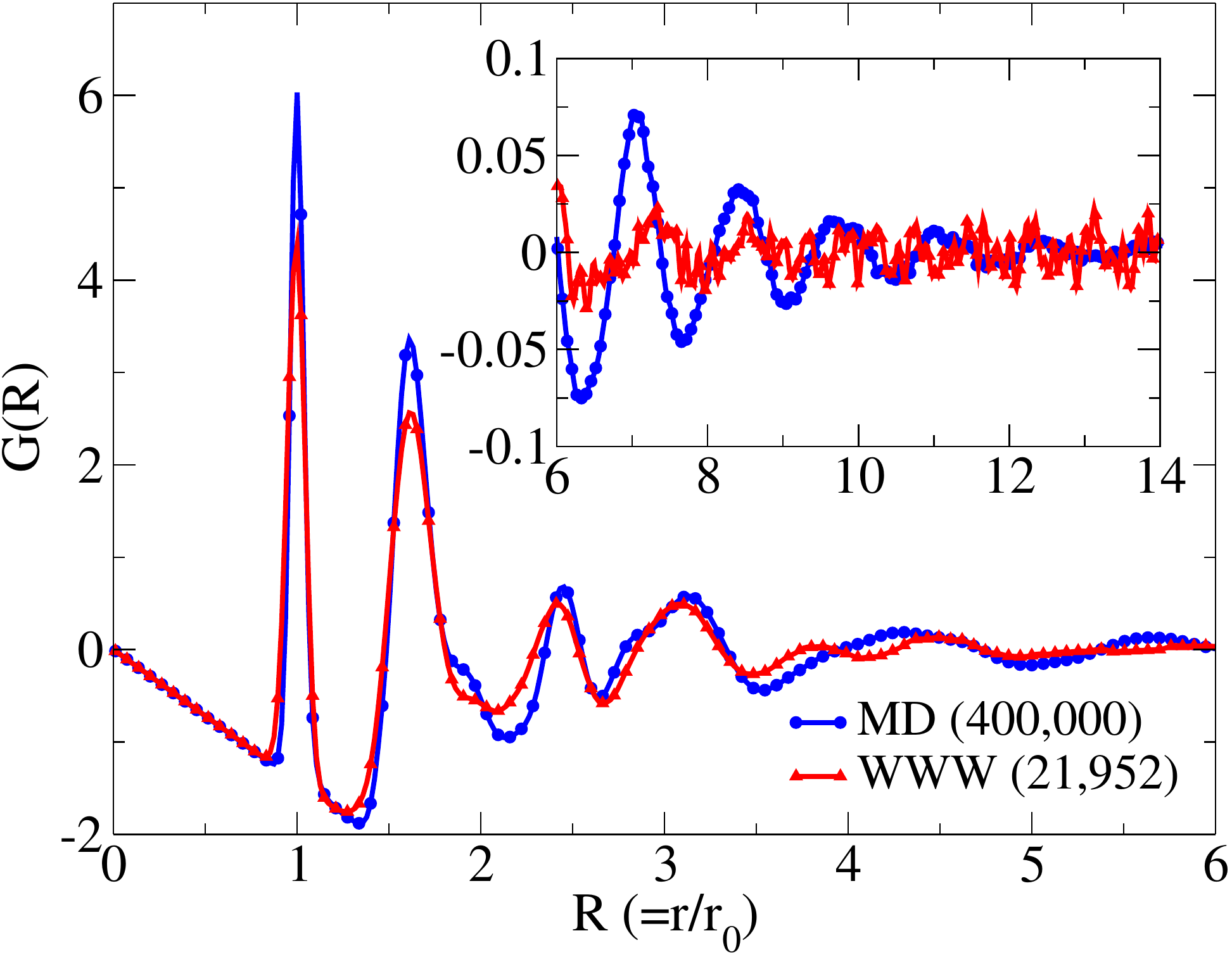}
\caption{
The reduced PCFs of the 400,000-atom MD models 
and 21,952-atom WWW models showing the presence 
of considerable extended-range oscillations in 
larger MD models. For clarity, the radial 
distances are scaled by the corresponding first 
peak of the PCF, i.e., $R = r/r_0$, where 
$r_0$=2.37 {\AA}.  
}
\label{400MD}
\end{figure}

Figure \ref{rdf_SI} shows the reduced PCF obtained 
from the MD models of {\asi}, which consist of 400,000 
atoms. The data presented here correspond 
to the configurational-averaged values of $G(r)$ 
from three independent configurations. The inset 
in Fig.\,\ref{rdf_SI} shows the presence of distinct 
radial oscillations at a distance beyond 20 {\AA}, 
extending at least up to 40 {\AA}. 
Similar oscillations have been also observed in 
the reduced PCF of 21,952-atom WWW models, but 
in a somewhat weaker form.  
This is apparent in Fig.\,\ref{400MD}, where we 
have plotted the configurationally averaged reduced 
PCFs for the 400,000-atom MD models and 21,952-atom 
WWW models.  For comparison, the radial distances 
($r$) in Fig.\,\ref{400MD} are scaled by the 
corresponding position of the first peak ($r_0$) 
by introducing a scaled variable $R = r/r_0$.  The 
inset in Fig.\,\ref{400MD} clearly shows the 
presence of considerable oscillations in larger 
400,000-atom MD models compared to their WWW 
counterpart in the region of $R$ from 6 to 14, which 
translates into a distance of 14 {\AA} to 33 {\AA} 
for $r_0 \approx$ 2.37 {\AA}. The observed differences 
can be partly attributed to the size and statistics 
and in part to the nature of simulations. In 
general, MD models are considered to be more 
representative of annealed samples of {\asi}, which 
are slighly more ordered than their as-deposited 
counterpart. 

Table \ref{tab1} presents some 
characteristic structural properties of the MD and 
WWW models. Since the presence of too many structural defects 
can affect the local density of the networks, 
and the radial correlations between atoms, it 
is necessary for the models to exhibit properties 
that are compliant with experimental observations. 
The presence of only a few dangling bonds 
(up to 1.3\%) and floating bonds (up to 1.2\%), 
as well as a small value of the root-mean-square width, 
$\Delta \theta$, about 9--10{\deg}, of the 
bond-angle distribution, confirms that the 
structural properties of these models are indeed 
consistent with actual samples of {\asi}.  

\begin{figure}[t!]
\centering
\includegraphics[width=0.45\textwidth]{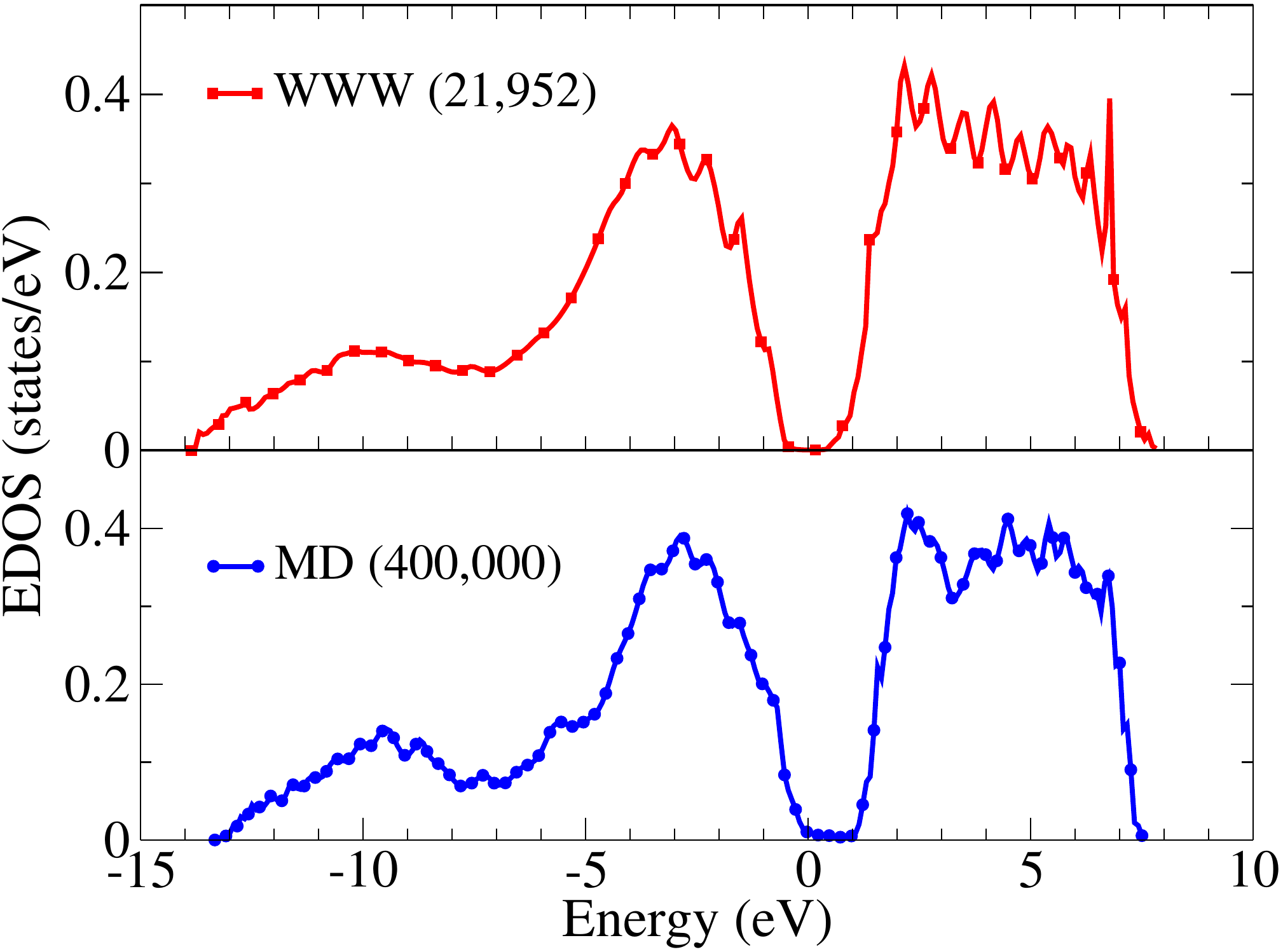}
\caption{
The electronic density of states (EDOS) 
of 21,952-atom WWW and 400,000-atom MD 
models obtained from the tight-binding 
approximation. The presence of an electronic 
band gap is clearly visible in the plots.
}
\label{EDOS}
\end{figure}

To further characterize the models, one often 
computes the electronic density of states (EDOS). 
The EDOS in {\asi} is found to be very 
sensitive to the presence of coordination 
defects, especially three-fold-coordinated Si 
atoms or dangling bonds. 
The presence of an electronic gap largely 
depends on these defects, and the size of 
the gap is known to be related to the 
density of such defects and the degree of 
disorder in bond-length and bond-angle 
distributions. We have therefore calculated 
the EDOS of 21,952-atom WWW models and 
400,000-atom MD models.  Since the diagonalization 
of the Hamiltonian matrix ($H$) of such 
large {\asi} models is highly nontrivial, we 
had to resort to: a) the tight-binding approximation 
of the Hamiltonian; and b) employ the recursion 
method of Haydock, Heine, and Kelly (HHK)~\cite{Haydock:1972,Haydock:1980} to 
obtain the EDOS. In the recursion approach of 
HHK, one calculates the projected density of states 
$n_{\alpha}(E)$, associated with a basis function 
$|\alpha\rangle$ (involving a site and an orbital), 
by writing 
\be
n_{\alpha}(E)= \sum_k |\langle \alpha|\psi_k\rangle|^2 \delta(E-E_k). 
\ee
Here, $E_k$ and $\psi_k$ are the 
energy eigenvalues and eigenvectors 
of $H$, respectively. Using a 
representation of the $\delta$-function 
and writing $z = E + \imath\epsilon$, where 
$\epsilon \to 0^{+}$,  it can be shown that the projected 
EDOS can be expressed in terms of the 
singular part of the diagonal element 
of the resolvent of $H$ or the Green's operator 
$ \Ghat(z) = (z\Ihat-\Hhat)^{-1}$. This 
yields~\cite{green}
\be
n_{\alpha}(E) = -\frac{1}{\pi} \lim_{\epsilon \to 0^+} \Im \, G_{\alpha\alpha}(E + \imath\epsilon). 
\label{LDOS} 
\ee
The local EDOS obtained from using Eq.~(\ref{LDOS}) is 
averaged over multiple sites to calculate the 
total EDOS. For 400,000-atom MD models, the problem is particularly 
difficult due to the handling and storage of large matrices and the 
computational cost associated with the calculation for 
all sites. In practice, a few clusters of several 
hundred atoms are found to suffice for configurational 
averaging. Using a fast matrix-vector multiplication scheme 
and a compressed representation of the 
sparse $H$ matrix, one can implement an 
order-$N$ algorithm for the calculation of the  
local EDOS in the tight-binding approximation. 
The results obtained from these calculations 
are shown in Fig.\,\ref{EDOS}. The presence 
of a clean gap, rather than a pseudo gap, in 
the EDOS further establishes 
the quality of the models. The approach can 
be adapted to calculate the vibrational density 
of states in the harmonic approximation, 
provided that an efficient scheme to obtain 
electronic forces for the construction of the 
dynamical matrix (DM) of {\asi} is available. 
An accurate order-$N$ approach to construct 
the DM within the framework of tight-binding 
formalism can be found in Ref.~\onlinecite{Biswas:2002}.

\begin{table}[t!]
\caption{
Structural properties of three WWW models (W1--W3) 
and three MD models (MD1--MD3). The average bond 
length ($\langle r \rangle$), average bond angle 
($\langle \theta \rangle$), and the root-mean-square 
width of bond angles ($\Delta \theta$) are expressed 
in {\AA} and degree, respectively. $C_n$ indicates 
the number of $n$-fold-coordinated atoms 
(in percent). \\
}
\centering 
\begin{tabular}{ c| c|c | c | c | c|c  |c|c  }
\hline
 \multicolumn{2}{c|} {Model}& \multicolumn{2}{c|} {Bond angle} &  \multicolumn{4}{c|} {Atomic coordination}& Bond length \\
 \cline{1-8}
Type & Size $(N)$ & $\langle \theta \rangle$ & $\Delta \theta$ & $C_{2}$&$C_{3}$ &$C_{4} $ & $C_{5} $&$\langle r \rangle$  \\
\hline
W1  & 21,952  &109.21  &10.04 & 0.00 & 0.00 & 99.86 & 0.14 & 2.36 \\
W2  & 21,952  &109.23  &9.83  & 0.00 & 0.00 & 99.9  & 0.1  & 2.36\\
W3  & 21,952  &109.22  &9.87  & 0.00 & 0.00 & 99.88 & 0.12 & 2.36\\
MD1 & 400,000 &109.23  &9.26  & 0.02 & 1.28 & 97.59 & 1.11 & 2.38 \\
MD2 & 400,000 &109.23  &9.31  & 0.03 & 1.29 & 97.52 & 1.16 & 2.38 \\
MD3 & 400,000 &109.23  &9.34  & 0.02 & 1.26 & 97.57 & 1.15 & 2.38 \\
\hline
\end{tabular}
\label{tab1}
\end{table}

\subsection{Origin of extended-range oscillations in {\asi}}

\begin{figure}[t!]
\centering
\includegraphics[width=0.5\textwidth]{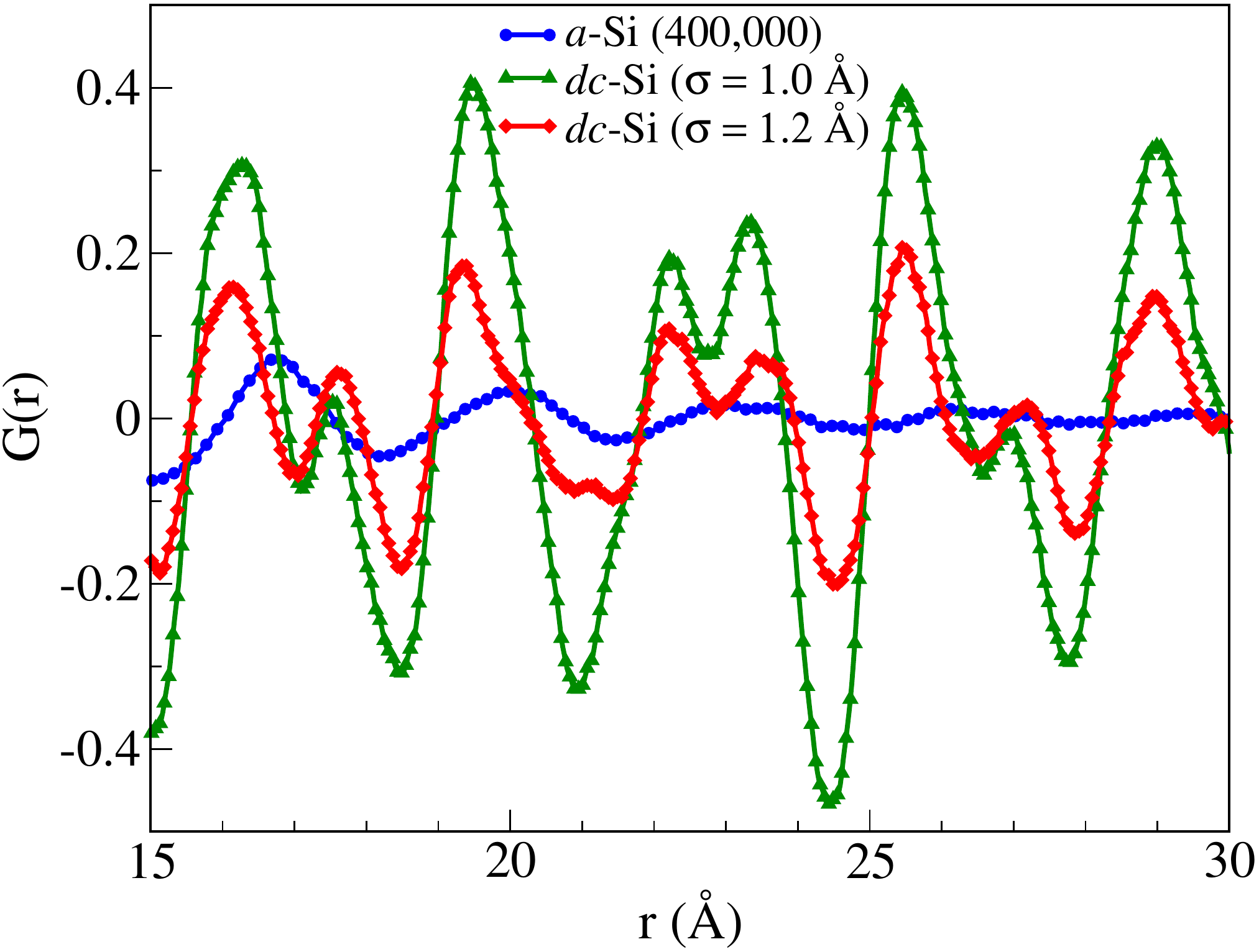}
\caption{
The presence of radial oscillations in MD models 
of {\asi} (blue) of size 400,000 atoms and two {\dcsi} 
structures of size 405,224 atoms from 15 {\AA} 
to 30 {\AA}. The positions of the radial peaks of 
{\asi} approximately correspond to those of {\dcsi}, 
indicating the possible presence of weak extended-range 
ordering in {\asi} beyond 15 {\AA}.
}
\label{rdf_CSI}
\end{figure}

The first step toward understanding the ERO in {\asi} 
follows from an analysis of the reduced PCF of 
disordered crystalline silicon ({\dcsi}) structures. 
The inclusion of positional disorder washes out the 
sharp ${\delta}$-functions in the PCF of diamond 
{\csi} and leads to a series of broadened peaks for 
the resulting {\dcsi} structures. A comparison of the 
reduced PCF of {\asi} with those from {\dcsi}, 
for $\sigma$ = 1.0 {\AA} and 1.2 {\AA}, in Fig.\,\ref{rdf_CSI} 
reveals that {\asi} exhibits small but noticeable 
oscillations at large distances 
of up to at least 30 {\AA}. Despite the fine structure of 
$G(r)$ in {\dcsi}, it is apparent that the positions
of the peaks in {\asi} approximately coincide with 
those in {\dcsi}. This observation leads to the 
possibility that the ERO in {\asi} 
could originate from the presence of weak radial-shell 
structures on the nanometer length scale, as in the 
case of {\dcsi}. This point is examined at length in 
the following paragraphs. 

\begin{figure}[t]
\centering	  
\includegraphics[width=0.5\textwidth]{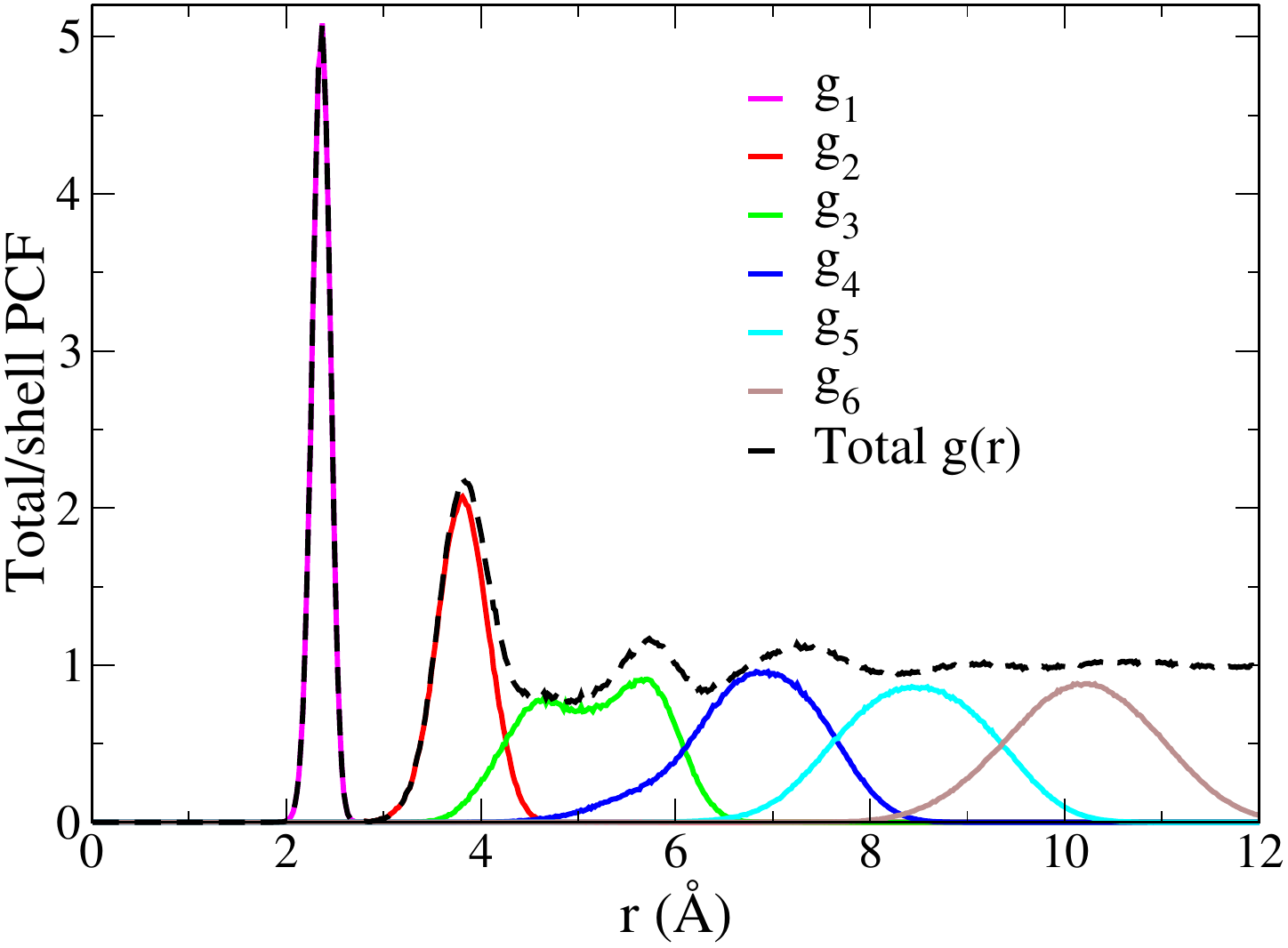}
\caption{
The shell pair-correlation function, $g_n(r)$,  
for the $n$th coordination shell of a 21,952-atom 
WWW model of {\asi}.  The total $g(r)$ (dashed 
black), which is given by the sum of all 
shell PCFs, is also shown in the plot.  
}
\label{shell}
\end{figure}

Assuming that radial-shell structures exist in the 
partially-ordered environment of {\asi} at large 
distances, one may express the total PCF, $g(r)$, 
as a linear combination of the same for each 
coordination shell, $g_n(r)$. Thus, $g(r) = \sum_n g_n(r)$, 
where $g_n(r)={\langle g(r=|\mathbf{r_n-R_i}|)\rangle}_i$. 
Here, $r$ is the distance between a central atom at 
$\mathbf R_i$ and its neighbors in the $n$th coordination 
shell at ${\mathbf r_n}$, and the symbol 
${\langle \, \rangle}_i$ stands for the average over all 
atoms and independent configurations. 
Since, for an arbitrary (highly) disordered network, distant radial 
shells may not exist or be well defined -- depending on the 
degree of radial disorder -- it is more appropriate to 
define the $n$th coordination or topological shell as 
one that consists of $n$th near neighbors of the central 
atom at $\mathbf{R_i}$. 
\begin{figure}[ht]
\centering	  
\includegraphics[width=0.3\textwidth]{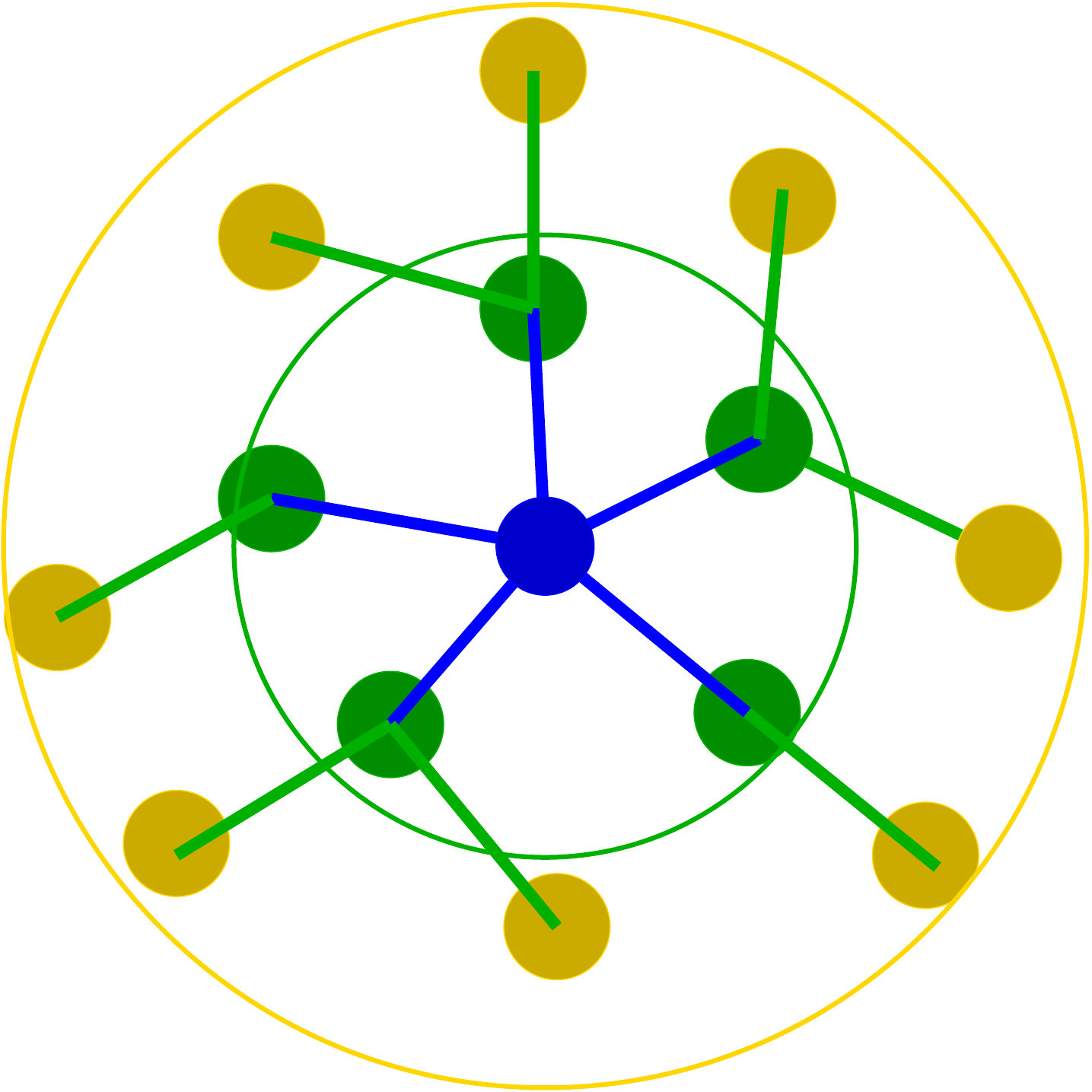}
\caption{
A schematic representation showing the first two 
coordination shells of a central atom (blue) in 
a two-dimensional disordered network. The atoms 
in the first shell (green) and the second shell 
(yellow) can be reached from the central 
atom in one step and two steps, respectively. 
}
\label{2D}
\end{figure}
This is schematically illustrated in Fig.\,\ref{2D} by 
showing the first-shell neighbors (green) and the 
second-shell neighbors (yellow) of the central atom (blue). 
The key point here is that the $n$th 
neighbors of a central atom are those that can be reached 
(from the center) by a minimum of $n$ distinct and 
irreversible steps, irrespective of the presence of 
well-defined radial shells or not.  Thus, the 
coordination shells defined above depend on the 
topology or connectivity of the atomic network, and 
the three-dimensional shape of the shells may not 
be necessarily spherical.  We shall see later that this 
can lead to a highly asymmetrical radial distribution of atoms within 
the coordination shells of partially-disordered networks. 
Figure \ref{shell} shows the shell PCFs, $g_n(r)$, 
obtained for the first six coordination shells, 
along with $g(r)$ for a 21,952-atom WWW model of {\asi}. 
It is apparent that the shell PCFs, for $n$ = 1 to $n$ = 6, 
can be represented by a bell-shaped curve in 
{\asi}, with the exception of $g_3(r)$ for 
which a bi-modal distribution is observed. 
The latter is consistent with the 
earlier study by Uhlherr and Elliott~\cite{Uhlherr:1994}. 
The bi-modal shape of $g_3(r)$ originates from the 
distribution of the end-to-end radial distances of 
a set of four neighboring atoms or quartets associated 
with dihedral angles in {\asi}.

\begin{figure}[t!]
\centering	  
\includegraphics[width=0.5\textwidth]{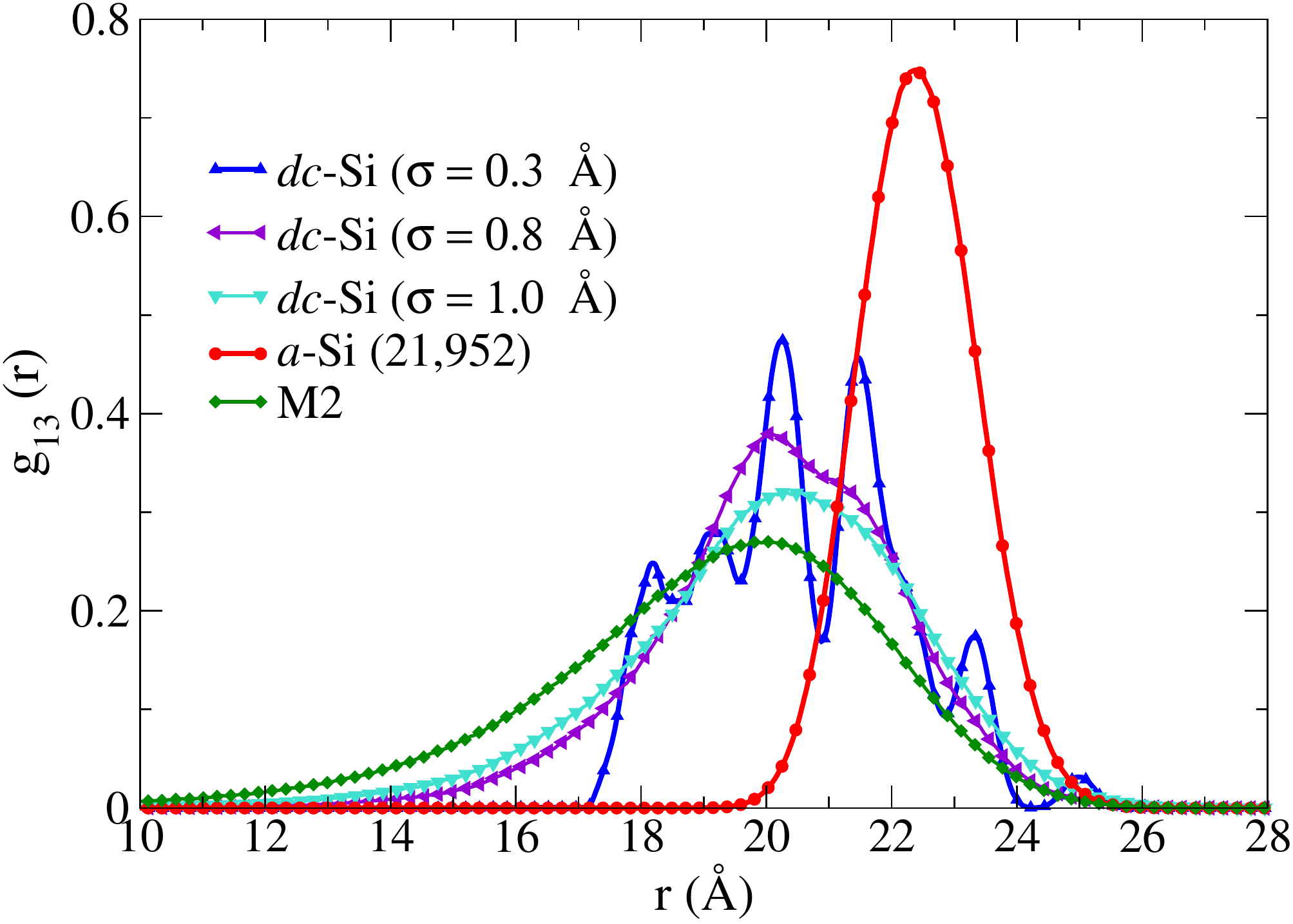}
\caption{
The shell pair-correlation functions, $g_{13}(r)$, 
obtained from 21,952-atom WWW models of {\asi} (red), 
{\dcsi} (blue/purple/cyan), and partially-ordered 
configurations M2 (green) of Si atoms. The disordered 
crystalline structures were generated from the 
diamond {\csi} structure, using $\sigma$ = 0.3, 
0.8, and 1.0 {\AA}. 
}
\label{g13}
\end{figure}

\begin{figure}[ht!]
\centering	  
\includegraphics[width=0.5\textwidth]{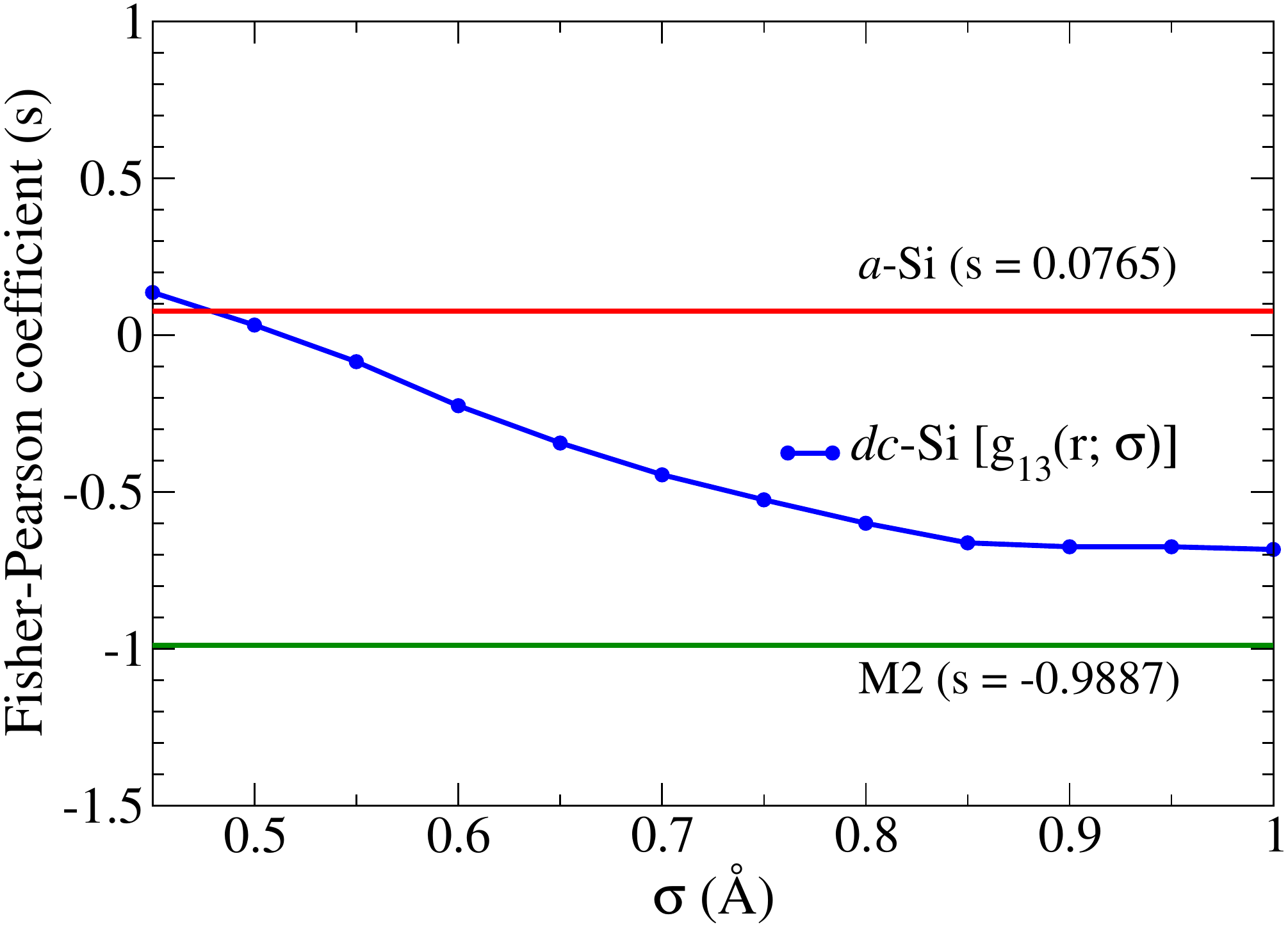}
\caption{
The variation of the Fisher-Pearson coefficient 
of skewness, $s$, for the distribution $g_{13}(r; \sigma)$ 
with $\sigma$ for a number of {\dcsi} models (blue). 
The coefficients for the M2 model (green) and 
{\asi} (red), for $\sigma$ = 0, are shown in the 
plot for comparison. 
}
\label{skew}
\end{figure}

Having expressed the total PCF in terms of $g_n(r)$, 
we now examine the oscillations in the PCF by 
studying individual $g_n(r)$s, which reflect the 
characteristic properties of the radial distributions 
of atoms in $n$th shells.  In particular, the width 
of $g_n(r)$ is indicative of the strength of the 
radial (dis)order in the $n$th shell. A small value of 
the width corresponds to a highly ordered state of 
atoms within the shell as far as radial ordering is 
concerned, and vice versa. This assertion can be 
verified by computing $g_n(r)$ for a number of 
partially-ordered networks of silicon.  Figure 
\ref{g13} shows the results for the 13th coordination 
shell, $g_{13}(r)$ as a representative example, 
obtained from 21,952-atom models, 
of {\asi}, {\dcsi}, and M2.  As stated earlier in Sect.\,II, 
the latter model (M2) is characterized by the presence 
of a well-defined first-coordination shell, 
whereas the {\dcsi} structures are produced by using 
a value of $\sigma$ in the range from 0.3 {\AA} 
to 1.0 {\AA}.  
It is apparent that a small value of $\sigma$ (for 
example, $\sigma$ = 0.3 {\AA}) produces well-defined 
multiple peaks in $g_{13}(r)$ for {\dcsi} models. However, 
as the value of $\sigma$ increases and goes beyond 0.6 {\AA}, 
the peaks in $g_{13}(r)$ coalesce to form a unimodal 
distribution. This is unsurprising due to the presence of 
strong residual crystalline order in the {\dcsi} networks 
for $\sigma \le$ 0.6 {\AA}. By contrast, the width of 
$g_{13}(r)$ for {\asi} is found to be considerably smaller 
than its M2 counterpart, which shows a more radially disordered 
distribution of atoms within the same shell in M2. This 
observation is found to be true not only for $g_{13}(r)$ 
but also for all $g_n(r)$s.
The high asymmetry of $g_{13}(r)$ for the M2 and {\dcsi} 
models can be readily attributed to the connectivity of 
the atoms in these models. Since the position of an atom 
in a given coordination shell is determined by the number 
of steps/hops from the central atom, there exist a few 
atoms in the shell that are radially close to the central 
atom but are not reachable (from the central atom) via a 
small number of steps/hops, due to the low connectivity 
of the atoms in the networks for increasing values of 
$\sigma$. This is reflected in the left tail of the 
distribution (see Fig.\,\ref{g13}), which leads to a 
negative value of the skewness for the radial distribution 
of atoms in the shell. This can be verified by computing 
the Fisher-Pearson (FP) coefficient of skewness~\cite{Fisher}, 
$s$, for $g_{13}(r)$, for different $\sigma$ values.  
In general, the FP coefficient of skewness is given 
by the standardized third central moment of a 
distribution, and a negative value of the coefficient 
signifies a skewed distribution toward the left, 
and vice versa. 
The variation of $s$ with $\sigma$ for a number of {\dcsi} 
models is shown in Fig.\,\ref{skew}, along with the 
corresponding value of the coefficient for the M2 
model for comparison.

\begin{figure}[t!]
\centering	  
\includegraphics[width=0.5\textwidth]{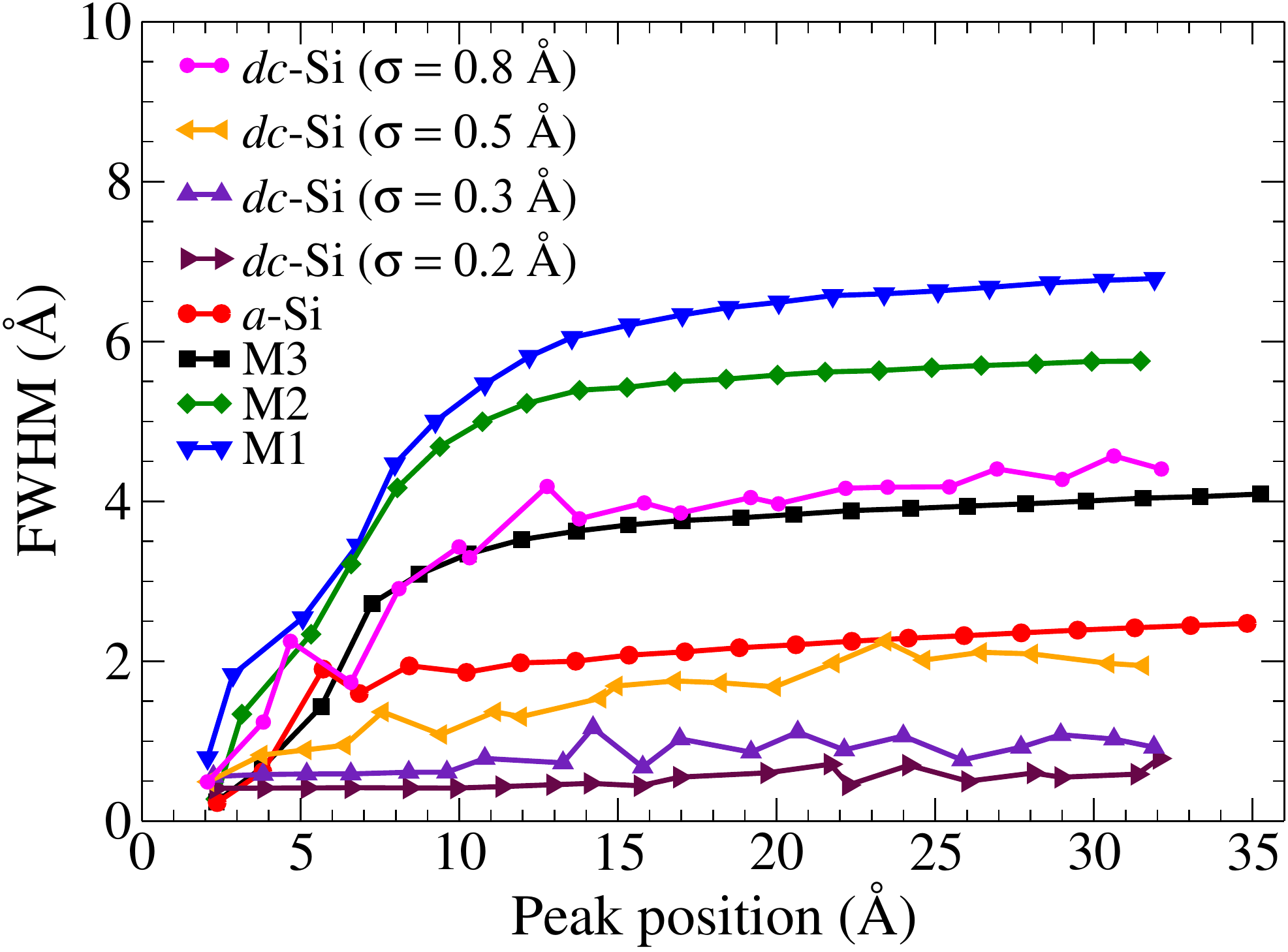}
\caption{
The full width at half maximum (FWHM) for a set of 
partially-ordered configurations (M1 to M3), {\asi}, 
and four {\dcsi} structures of size 21,952 atoms. 
The {\dcsi} structure (with $\sigma$ = 0.2 {\AA}) 
is the most ordered configuration, whereas M1 is 
the least ordered configuration, by construction.
}
\label{width}
\end{figure}

Figure \ref{width} shows the full width at half maximum 
(FWHM) of the shell PCFs, $g_n(r)$, for different 
shells, from $n$=1 to $n$=20, for a class of partially-ordered 
models (M1 to M3), {\asi}, and {\dcsi} models of size 21,952 
atoms. Since {\dcsi} models tend to exhibit the presence of 
multiple peaks in $g_n(r)$ for $\sigma \le $ 0.6 {\AA}, 
the FWHM for the {\dcsi} models (with multiple peaks) in 
Fig.\,\ref{width} is calculated by fitting each individual 
peak with a Gaussian distribution and averaging over the 
resulting FWHM values for all major peaks in the distribution. 
The FWHM values (in Fig.\,\ref{width}) suggest that 
the M1 models are highly disordered, whereas the {\dcsi} 
structures with $\sigma$ = 0.2 {\AA} are the least 
disordered configurations. This observation is 
indeed true by construction. For $\sigma$ = 0.2--0.5 
{\AA}, a significant radial ordering exists in the 
{\dcsi} structures that leads to a small value of 
the width in Fig.\,\ref{width}. The rest of the 
models, from M2 and M3 to {\asi}, exhibit an 
increasingly more ordered state of radially distributed 
atoms in the shells. It is apparent that as more 
radial ordering is incorporated in a model 
(for example, M2 and M3), the corresponding FWHM 
value of $g_n(r)$ begins to decrease for a given 
shell. Conversely, the inclusion of (additional) structural 
disorder increases the corresponding FWHM value 
of $g_n(r)$ in a model. This can be seen from 
Fig.\,\ref{d-width1}, where the addition of positional 
disorder, via random displacements of atoms in M1, M3, 
and {\asi}, resulted in an increase of 
the FWHM values of $g_n(r)$. This observation also 
applies to the total PCF of {\asi}. Figure \ref{d-width2} 
shows that the amplitude of the radial oscillations 
reduces in the region of 20--40 {\AA} with the 
addition of positional disorder in {\asi}. 
It may be noted that the FWHM values for the M1 models, 
which are highly disordered by construction, are 
practically unaffected in Fig.\,\ref{d-width1} 
in the presence of additional disorder with $\sigma$ 
values of the order of 0.3 {\AA}.
Thus, the width (or the average width for a multimodal 
case) of $g_n(r)$ can be taken as a measure of the 
radial order/disorder in partially-ordered networks, 
including {\asi} and {\dcsi} structures.

\subsection{Shannon information as a measure of 
extended-range ordering} 

\begin{figure}[t!]
\centering	  
\includegraphics[width=0.5\textwidth]{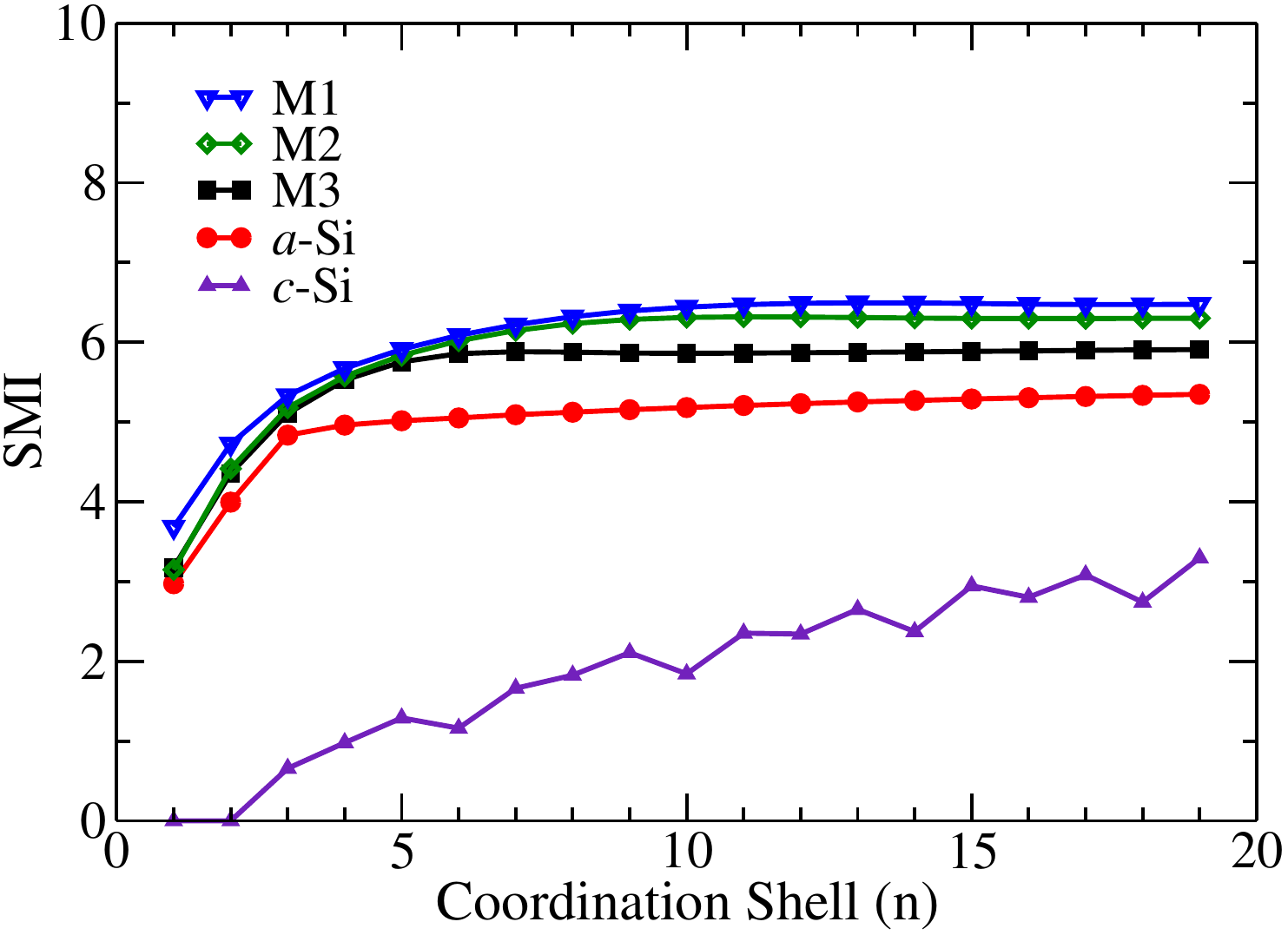}
\caption{
The Shannon measure of information (SMI), 
associated with a discrete probability measure 
$p_n$, obtained from the shell pair-correlation 
functions, $g_n(r)$, for M1 to M3, {\asi}, and 
diamond {\csi}. For disordered and amorphous 
Si networks, the results were averaged over 
three independent configurations for each shell. 
}
\label{SMI}
\end{figure}

The assertion that the width of the shell pair-correlation 
function, $g_n(r)$, can provide a measure of the 
disorder in the radial distribution 
of atoms in the $n$-th coordination shell of a disordered 
network is not particularly surprising and it directly 
follows from the Shannon measure of information 
(SMI)~\cite{Shannon:1948}. By normalizing the 
shell PCF, $g_n(r)$, one can readily construct a 
discrete probability measure, $p_n^i$, to define 
the SMI
\be 
S[p_n^i] = -k\, \sum_i p_n^i \, \ln p_n^i. 
\label{eq2} 
\ee 
In Eq.(\ref{eq2}), the value of $0\ln(0)$ is 
defined to be 0, $k$ is a constant, and $p_n^i$ 
is given by 
\[ 
p_n^i = \frac{g_n(r_i)}{\sum_i g_n(r_i)}. 
\] 
The SMI can be understood as providing a measure 
of the degree of uncertainty or the lack of 
radial ordering in the distribution of atoms in 
the coordination shells. The multiplicative 
constant $k$ in Eq.\,(\ref{eq2}) can be taken 
as unity without any loss of generality. 
The results for the SMI obtained from M1/M2/M3/{\asi} 
models are shown in Fig.\,\ref{SMI} for the first 
twenty coordination shells. The corresponding results 
for diamond {\csi} are also shown in the plot 
for comparison. As one may expect, the SMI values for 
different shells behave in a similar manner as that 
of the FWHM (of the shell PCFs) with respect to 
the peak position in Fig.\,\ref{width}. Once again, the largest 
values of the SMI correspond to the highly 
disordered M1 models, whereas {\asi} exhibits the 
smallest values of the SMI for each shell among M1, M2, 
M3 and {\asi}. It is noteworthy that, unlike the case of 
disordered and amorphous Si networks,  the SMI values 
associated with the coordination shells in the 
diamond {\csi} structure, which is perfectly ordered, increase
considerably with the increasing shell number in a 
global sense. This observation can be attributed to 
the presence of multiple peaks in 
the higher-order coordination shells. Since the (shell) 
PCFs for a crystalline structure consist of a series 
of $\delta$-functions, the presence of an increasing 
number of peaks in the distant 
shells leads to more uncertainty in the radial distribution 
of the atoms in these shells.  This 
is reflected in the larger value of the SMI for 
the distant shells. Thus, the SMI can be loosely 
interpreted as a global measure of ordering/disordering in 
the distribution, which is most appropriate for 
describing the degree of order/disorder associated 
with unimodal distributions. 
However, for multimodal distributions, such 
as the {\dcsi} structures with $\sigma \le$ 0.5 {\AA}, 
one requires a suitable local measure of information, 
for example, the Fisher information~\cite{Fisher:1959}, 
in order to quantify the degree of disorder or uncertainty 
associated with the radial distribution of the atoms in the 
coordination shells. 
These issues will be addressed elsewhere from 
an information-theoretic point of view in a 
future communication.

\begin{figure}[t!]
\centering	  
\includegraphics[width=0.5\textwidth]{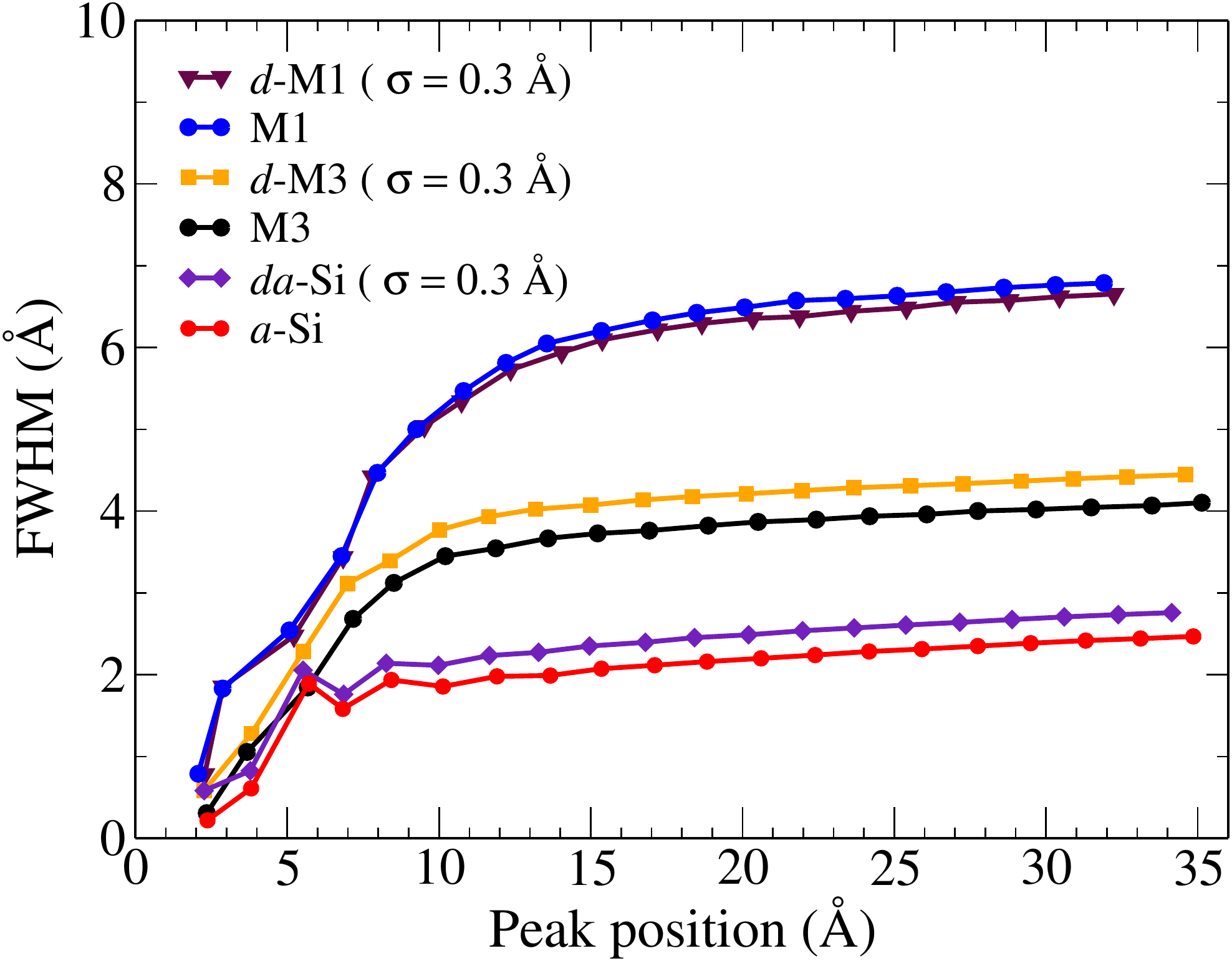}
\caption{
The effect of the addition of positional disorder 
on the FWHM of the shell PCFs in M1, M3, and {\asi}. 
The FWHM values of the unperturbed models are 
also shown for comparison. 
}
\label{d-width1}
\end{figure} 

\begin{figure}[t!] 
\centering
\includegraphics[width=0.5\textwidth]{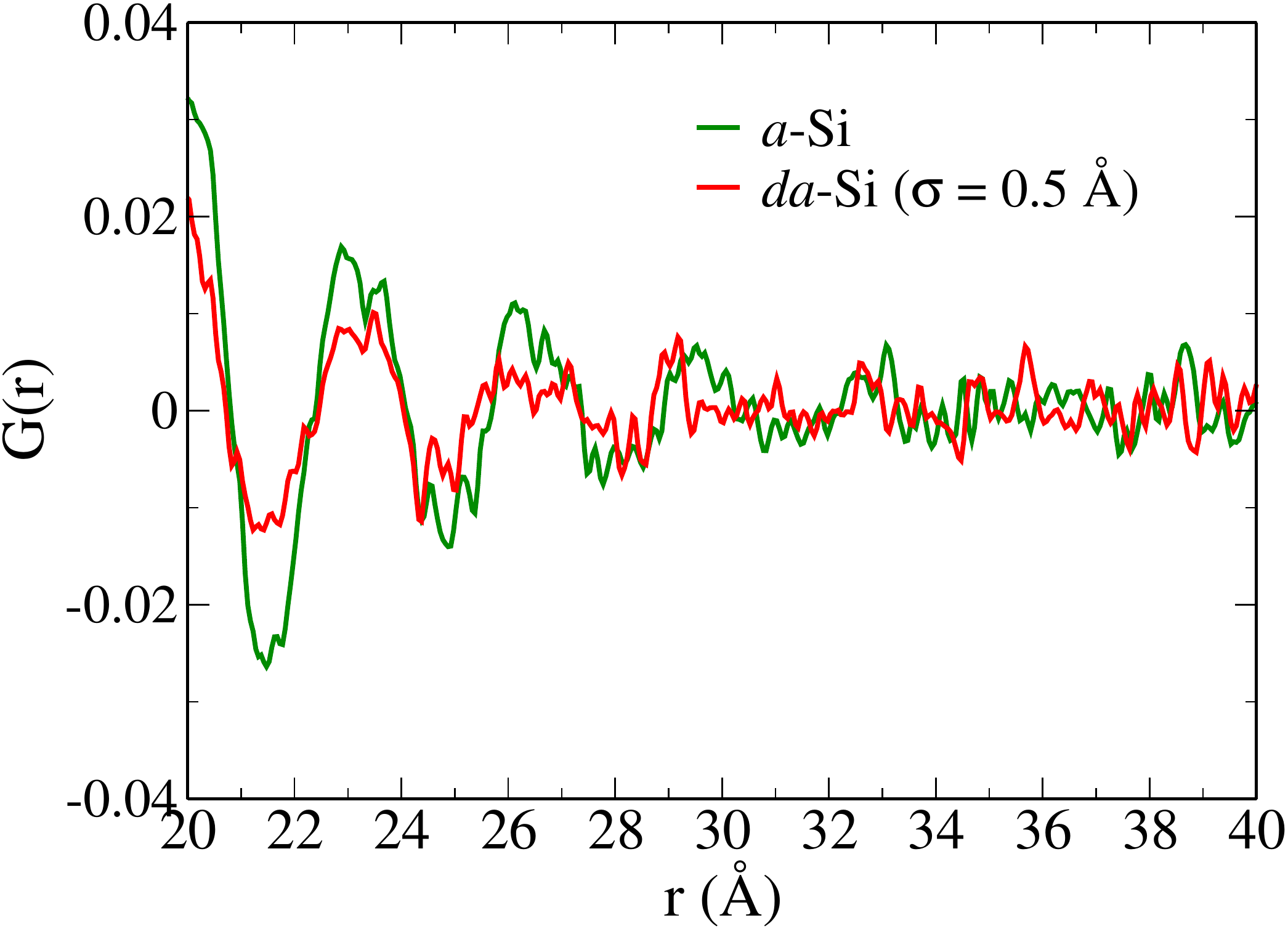} 
\caption{
The effect of the addition of positional disorder, with 
$\sigma$ = 0.5 {\AA}, on the radial oscillations in a 
400,000-atom MD model of {\asi} between 20 {\AA} and 
40 {\AA}.  The results for the corresponding pristine 
{\asi} model (green) are also shown for comparison. 
}

\label{d-width2}
\end {figure}

The origin of the extended-range oscillations in {\asi} 
can now be interpreted in light of the results from 
Figs.\,\ref{shell}--\ref{SMI}. Since the full PCF 
can be expressed in terms of its partial components, 
any structural aspects of $g(r)$, such as the 
extended-range oscillations, can also be represented 
by a suitable set of $g_n(r)$, associated with the 
length scale of the oscillations. Figures 
\ref{width} and \ref{SMI} essentially suggest that, 
as the degree of radial ordering in the full PCF 
increases from M1 to M3, the corresponding width 
and the Shannon information associated with $g_n(r)$ 
steadily decrease. 
Thus, the inclusion of radial information of up to a 
distance of 4 {\AA} in M2 and about 6 {\AA} in M3 
suffices to result in a reduction of the width of 
$g_n(r)$ associated with the distant coordination 
shells. Since {\asi} is characterized by the presence 
of strong radial ordering at least up to a length 
of 20 {\AA} in the full PCF, it is unsurprising 
that a small value of the width of $g_n(r)$ of 
{\asi} is reflective of the radial ordering in 
the distant shells on the length scale of 20--40 {\AA}.  
By contrast, the {\dcsi} models with $\sigma$ = 0.2--0.5 
{\AA} show significant radial ordering as far as 
the widths of various $g_n(r)$s are concerned.  
Thus, the ERO in {\asi} can be understood as the 
resultant density fluctuations, originating from 
highly ordered radial distributions of atoms in 
the first few coordination/radial shells, which propagate 
and decay radially as the (density) fluctuations 
travel through the distant shells.  A comparison 
of the results from the M2, M3, and {\asi} models 
in Fig.\,\ref{width} appears to suggest that 
the characteristic local radial ordering of up to 
6 {\AA} forces the atoms in distant shells to 
organize in such a way that small radial oscillations 
are built up on the length scale of up to 40 {\AA}, 
when the model is sufficiently large.  
The presence of these small but distinct radial 
oscillations in the full PCF is indicative of 
the existence of weak extended-range radial 
ordering in {\asi} up to a length of 40 {\AA}, 
as far as the size of the {\asi} models studied 
in this work are concerned.

\begin{figure}[t!] 
\centering
\includegraphics[width=0.5\textwidth]{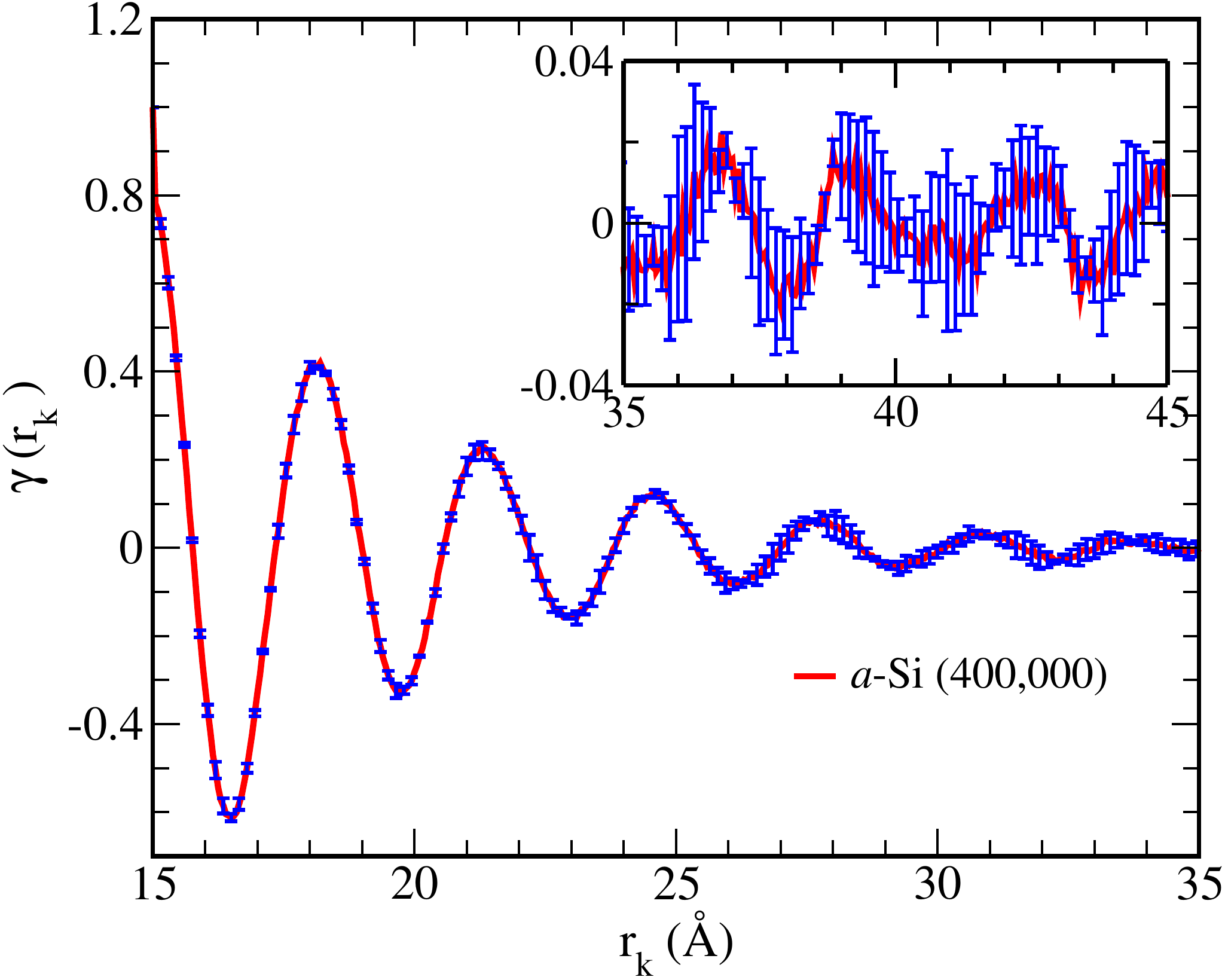} 
\caption{
The autocorrelation coefficient, $\gamma(r_k)$, 
of a set of $G(r)$ values from $r$ = 15 {\AA} to 
$r$ = 45 {\AA}, constructed from 400,000-atom 
MD models of {\asi}, showing the presence of 
radial correlations up to 45 {\AA}. The root-mean-square 
fluctuations are shown as error bars (blue vertical 
lines). For visual clarity, the results for the radial 
region from 35 {\AA} to 45 {\AA} are shown in 
the inset. 
}
\label{coreal}
\end {figure}

\subsection{
Decay of radial correlations, autocorrelation
coefficient, and comparison with experimental
diffraction data
}

The presence of radial atomic correlations 
beyond 20 {\AA} can be further evidenced by 
computing the autocorrelation coefficient(s) 
of $G(r)$. Assuming that $M$ observations, 
$y_1$, $y_2$, \ldots, $y_M$, form a time series, 
where $y_i = G(r_i)$, the autocovariance 
coefficient~\cite{cov}, $c_k$, between the 
observations that are $k$-steps apart, is 
given by: 
\be
c_k= \frac{1}{M} \sum_{i = 1}^{M-k} (y_i - \bar{y}) (y_{i+k} - \bar{y}), 
\: k = 1,...,n, n < M. 
\label{eq3} 
\ee
The autocorrelation coefficient at lag $k$ is then expressed as 
$\gamma_k = c_k/c_0$, where $c_0$ is the variance and 
${\bar y}$ is the mean value of the set $\{y_i\}$. 
Figure \ref{coreal} shows a plot of $\gamma_k$ 
versus $r_k$. 
Here, the set $\{y_i\}$ is constructed by choosing a 
segment of $ G(r)$ from $r_1$ = 15 {\AA} to $r_n$ = 45 {\AA} 
and expressing $k$ in terms of $r_k = r_1 + k\Delta r$, where 
$\Delta r$ is the distance between two consecutive observations 
of $G(r)$.  
It is apparent from the plot that, given the set of $G(r)$ values 
from 15 {\AA} to 45 {\AA}, $\Delta r$ = 0.05 {\AA}, 
$n$ = 600, and $M$ = 900, the radial correlations decay in an 
oscillatory manner and become almost negligible after 
35 {\AA}. The root-mean-square (RMS) fluctuations of $\gamma_k$, 
obtained from the configurational averaging of the results 
from three independent MD models of size 400,000 atoms, are 
also shown in Fig.\,\ref{coreal}. Since the RMS values of the 
fluctuations are almost of the order of $\gamma_k $ for 
$ r \ge $ 40 {\AA}, the radial correlations in this region may 
not be significant, even though the presence of small 
residual correlations can be seen in this region.

\begin{figure}[t!] 
\centering
\includegraphics[width=0.5\textwidth]{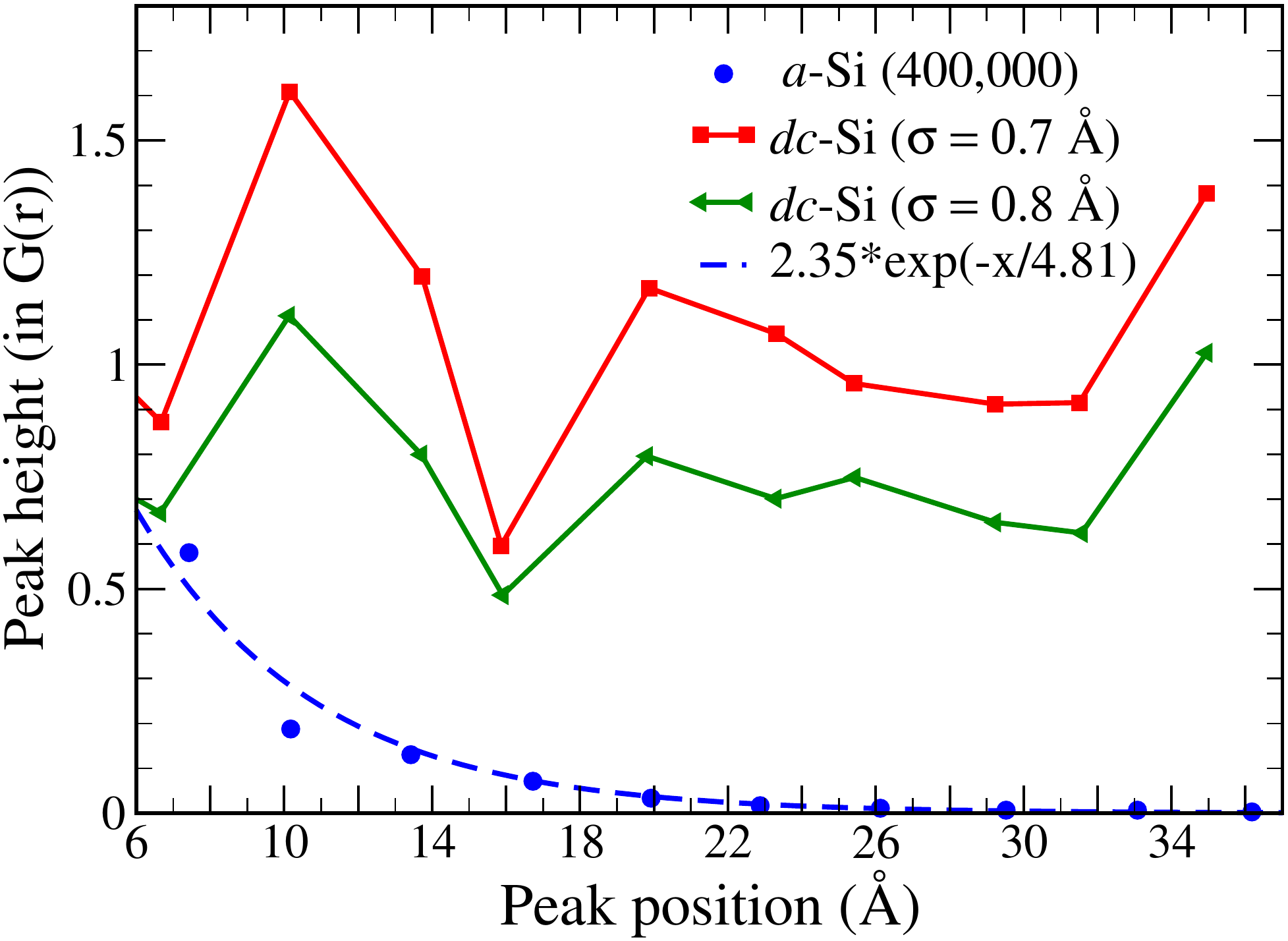}
\caption{
The decay of the radial peak heights in $G(r)$ for {\asi} 
(blue) and {\dcsi} (red and green) with 
peak distances. The exponential fit of the peak 
positions (dashed blue line) corresponds to the 
data for {\asi} models of size 400,000 atoms. 
The decay length for {\asi} corresponds to a 
value of 4.81 {\AA}. The corresponding peak positions 
for {\dcsi} structures are shown for comparison.  
}
\label{decay}
\end {figure}

We now provide a direct comparison of our results 
with those from diffraction measurements by computing the 
decay length and the (spatial) period of the 
ERO at radial distances beyond 10 {\AA}. 
High-energy x-ray diffraction measurements 
on {\asi} samples, by Roorda et al.~\cite{Roorda:2012}, 
suggest that the period of oscillations ranges 
from 2.77 {\AA} to 3.03 {\AA} and that the 
decay length in annealed samples of {\asi} 
is about 4.23 {\AA}.  
Figure \ref{decay} shows the decay of the amplitudes of 
radial oscillations in $G(r)$ for 400,000-atom MD models 
of {\asi}, which can be roughly considered as the 
simulated counterpart of annealed samples of {\asi} 
in experiments. Here, the amplitudes and positions 
of the peaks are obtained from Fig.\,\ref{rdf_SI}. 
The corresponding decay for {\dcsi} 
networks for $\sigma$ = 0.7 {\AA} and 0.8 {\AA} are 
also included in the plot for comparison. For visual 
clarity, the first three peaks of {\asi} are omitted 
from the plot, by choosing an appropriate range for 
the $y$ axis. The values of the period and the decay 
length obtained from our calculations compare very 
well with the results from experiments. 
The average period of oscillations from 400,000-atom 
MD models, in Fig.\,\ref{decay} (and Fig.\,\ref{rdf_SI}), 
is found to be 3.2$\pm$0.065 {\AA}, which is very 
close to the experimental value of 3.03 {\AA}, and 
the corresponding decay length turns out to be about 
4.81$\pm$0.012 {\AA}. The latter is somewhat higher 
than the experimental value of 4.23 {\AA}, obtained 
from the Fourier transform of experimental diffraction 
data by~\citet{Roorda:2012}. It is evident from 
Fig.\,\ref{decay} that the {\dcsi} models exhibit 
a rather slow decay, even for considerably large 
values of $\sigma$ from 0.7 {\AA} to 0.8 {\AA}. 

\subsection{Relation between ERO and the first sharp 
diffraction peak in {\asi}}

In this section, we address the question whether the 
presence of extended-range oscillations has any 
bearing on the position and intensity of the first sharp diffraction 
peak (FSDP) in {\asi}. 
\begin{figure}[t!]
\centering
\includegraphics[width=0.5\textwidth]{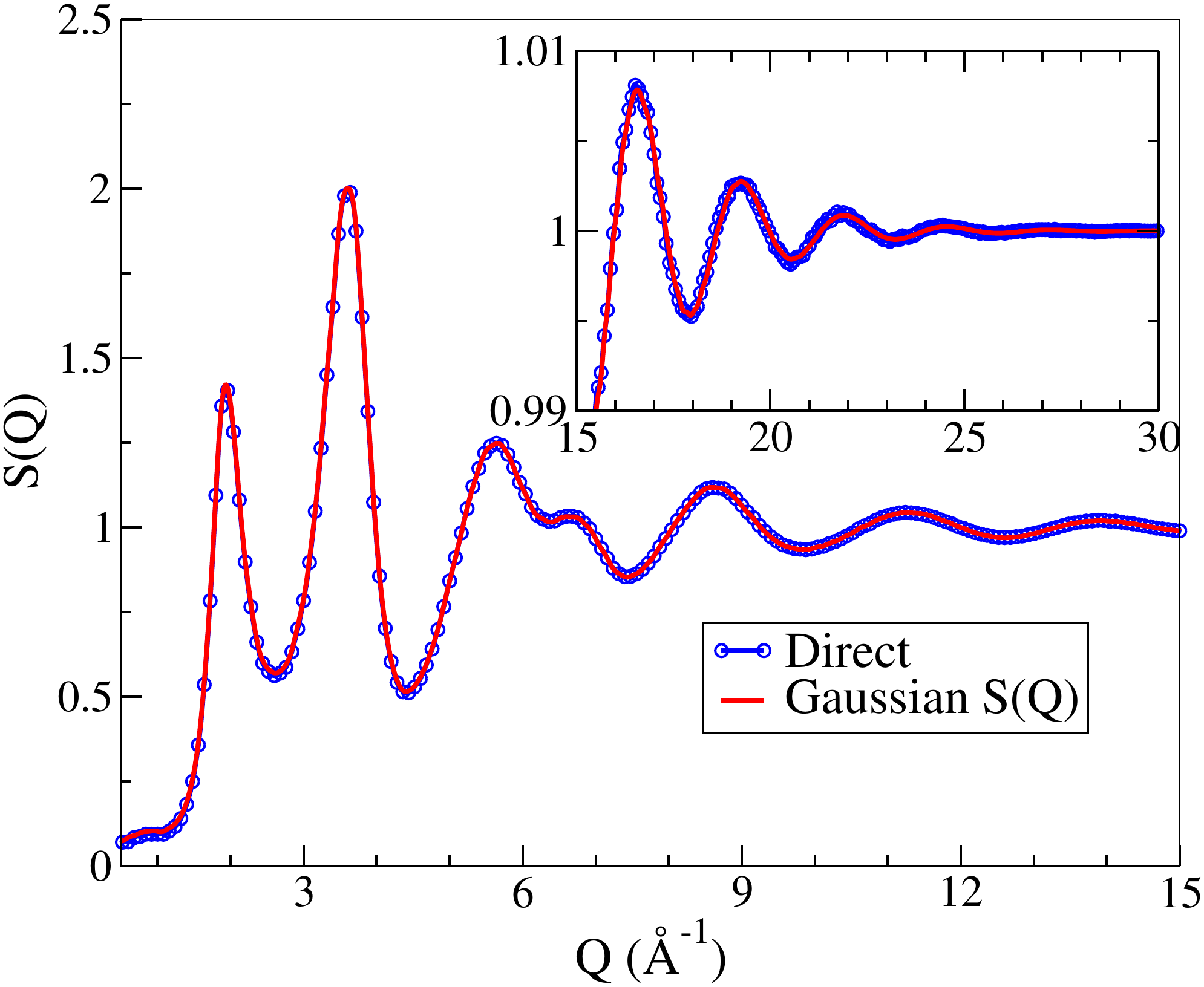}
\caption{
The static structure factor, $S(Q)$, of a 21,952-atom WWW 
model of {\asi} obtained from the Gaussian approximation 
(red line) and from direct numerical calculations 
(blue circles) using Eq.\,(\ref{eq5}). 
}
\label{SQ-approx}
\end{figure}
Since the origin of the FSDP is 
strongly related to the presence of medium-range order 
(MRO) in glasses, which can extend up to a radial 
distance of approximately 20 {\AA}, it is instructive 
to examine whether the ERO in {\asi} can produce any 
observable effect on the intensity of the FSDP near 
2.0 {\AA}$^{-1}$. Noting that the oscillations are particularly pronounced 
in $G(r)$, it is useful to write $G(r)$ as a linear 
combination of Gaussian functions,
\be
G(r) = \sum_{i=1}^m a_i e^{-b_i (r - c_i)^2}, 
\label{eq4}
\ee
in an effort to obtain an analytical expression 
for $S(Q)$ in terms of the Gaussian parameters.
The parameters $a_i$, $b_i$, and $ c_i$ determine the 
approximate peak/trough height, width, and the (radial) 
position of the $i^{\text{th}}$ peak/trough, respectively, 
and can be obtained either via a nonlinear fit of 
Eq.\,(\ref{eq4}) to experimental/simulated reduced 
PCF data, $G(r)$, or by minimizing a suitable cost 
function with respect to the set of parameters $(a_i, 
b_i, c_i)$. Here, we have taken the second approach 
and ensured that $b_i > 0$ for all $i$. The structure 
factor can be expressed in terms of the fitted Gaussian 
parameters: 
\bea
S(Q) & = & 1 + \int_0^{R_c} rG(r) \,\frac{\sin(Qr)}{Qr} \, dr  \label{eq5} \\
     & = & 1 + \frac{1}{Q} \sum_{i=1}^{m} a_i\sqrt{\frac{\pi}{b_i}}\, \sin(Qc_i)
\, \exp\left[-\frac{Q^2}{4b_i}\right]
\label{eq6}
\eea
\noindent 

\begin{figure}[t!] 
\centering
\includegraphics[width=0.5\textwidth]{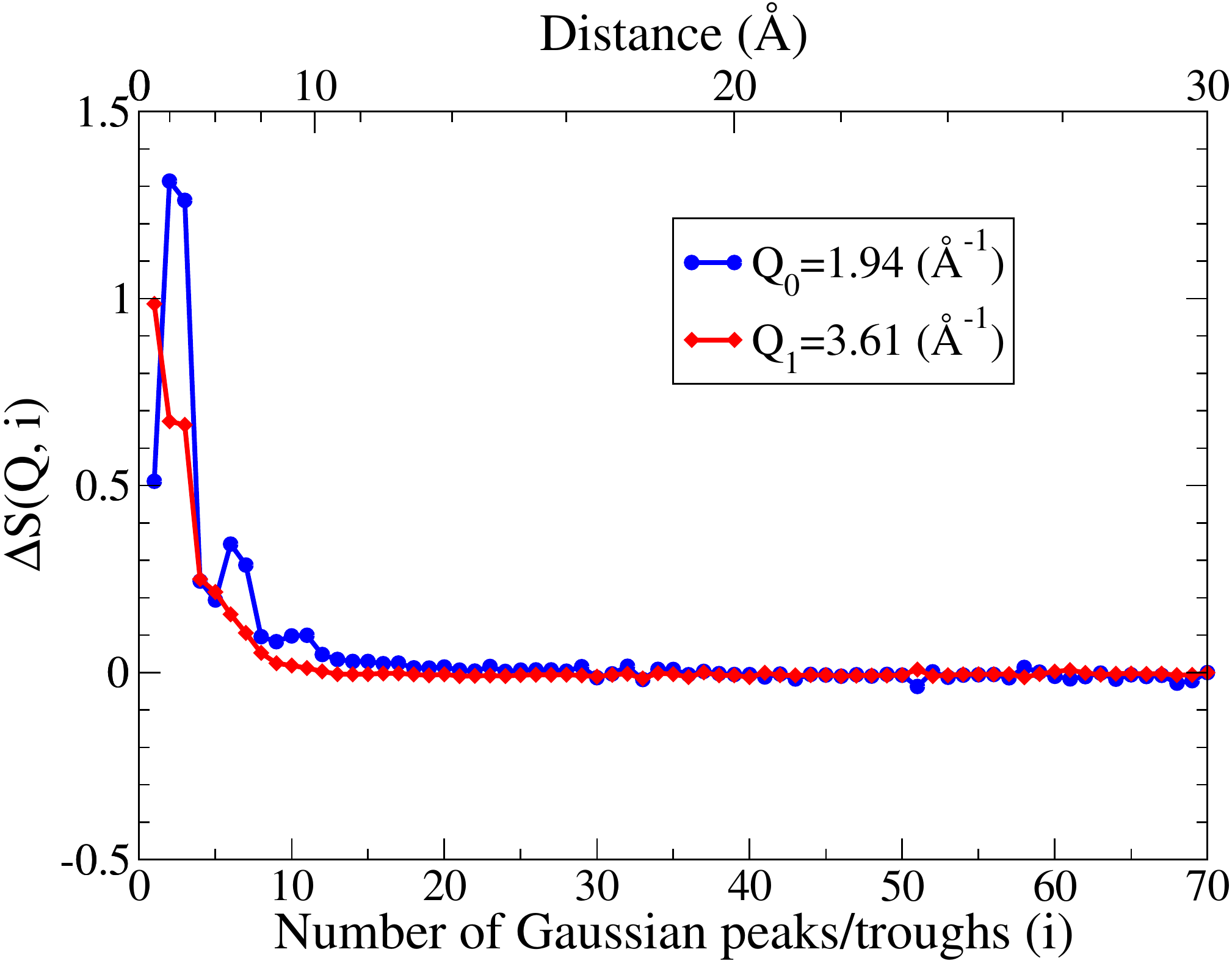}
\caption{
The convergence of the intensity of the FSDP (at $Q_0$) 
and the principal peak (at $Q_1$), obtained from 
Eq.\,(\ref{eq5}), with respect to the number ($i$) 
of Gaussian peaks/troughs for a 21,952-atom WWW model. 
The radial length associated with the Gaussian 
peaks/troughs is indicated in {\AA} on the secondary 
$x$ axis (top). 
}
\label{conv}
\end {figure}

In writing Eq.\,(\ref{eq6}), we have denoted, for notational
convenience, the set $(a_i, b_i, c_i)$ as the fitted
values of the parameters and assumed that the center of
each Gaussian function, $c_i$, satisfies the condition
$0 \ll c_i \ll R_c$ so that $S(Q)$ can be written as a 
sum of Gaussian integrals (and not error functions) with 
the integration limit extending from 0 to $\infty$. 
This condition is readily satisfied by choosing an 
appropriate value of $m$, such that $R_c \gg c_m$, 
and noting that the first peak of the PCF in {\asi} 
rapidly decays to zero for $r \le 2.0$ {\AA}. 
In practical calculations, a value of $R_c$ of the order 
of 20 {\AA} is found to be sufficient for accurate 
determination of $S(Q)$ using Eq.\,(\ref{eq5}) 
[see \citet{Dahal:2021} and Fig.\,\ref{conv} here].
The structure factor obtained from 
Eq.\,(\ref{eq6}) for a 21,952-atom WWW model of {\asi} is 
plotted in Fig.\,\,\ref{SQ-approx}, along with the results 
from direct numerical calculations from Eq.\,(\ref{eq5}) 
for comparison. For clarity, the wavevector region from 
15 {\AA}$^{-1}$ to 30 {\AA}$^{-1}$ is shown separately 
as an inset in Fig.\,\ref{SQ-approx}. 

The variation of the intensity of the FSDP 
and the principal peak (i.e., the peak at 
3.6 {\AA}$^{-1}$) can be studied, by using 
Eq.\,(\ref{eq6}), with respect to the number of Gaussian basis 
functions $m$ for a given $R_c$.  Writing 
$\Delta S(Q,i)=S(Q,m)-S(Q,i)$, where $m$=70 
for $R_c$ = 30 {\AA}, Fig.\,\ref{conv} shows 
the convergence of $\Delta S$ at $Q_0$ = 1.94 
{\AA}$^{-1}$ and $Q_1$ = 3.6 {\AA}$^{-1}$ for 
an increasing number ($i$) of peaks/troughs. 
Here, $Q_0$ and $Q_1$ correspond to the position 
of the FSDP and the principal peak, respectively. 
It is apparent that both $S(Q_0,i)$ and $S(Q_1,i)$ 
converge to the respective limiting value, $S(Q, m)$, 
very rapidly as $i$ approaches to 30, which corresponds 
to a radial length of about 18 {\AA}, as indicated 
in Fig.\,\ref{conv}. The length is indicated at 
the top of the plot as a secondary $x$ axis, 
which reflects the non-uniform distribution of 
Gaussian peaks/troughs in the radial-region of 
0--30 {\AA}. Figure \ref{conv} suggests that 
radial correlations from the region beyond 20 {\AA} 
do not really play any significant role. This 
observation can be stated more precisely. The 
magnitude of the contribution to $S(Q)$ obtained 
by including an additional peak/trough beyond 
$m$ in Eq.\,(\ref{eq6}) can be written as: 
\bea 
|\delta S(Q,m)| &=& |S(Q,m+1)-S(Q,m)| \notag \\
&=& \left| \frac{1}{Q} \sqrt{\frac{\pi a_{m+1}^2}{b_{m+1}}}\, 
\sin(Q c_{m+1}) \, \exp\left[-\frac{Q^2}{4b_{m+1}}\right]\right | \notag \\ 
&\le& 
\frac{1}{Q} \sqrt{\frac{\pi a_{m+1}^2}{b_{m+1}}} \, \exp\left[-\frac{Q^2}{4b_{m+1}}\right]. 
\label{eq7} 
\eea
Substituting $Q$ = $Q_0$ = 2 {\AA}$^{-1}$ in Eq.\,(\ref{eq7}) 
for the FSDP in {\asi}, one obtains: 
\be
|\delta S(Q_0, m)| < \frac{a_{m+1}}{\sqrt{b_{m+1}}} \, 
\exp\left[-\frac{1}{b_{m+1}}\right] 
\label{eq8} 
\ee
The asymptotic behavior of $|\delta S(Q_0,m)|$ with respect to $m$ 
follows from Eq.\,(\ref{eq8}). Since $G(r) \rightarrow$ 0 as $r 
\rightarrow R_c$ for very large models, the parameter $a_m$, which 
determines the height of the Gaussian peak, decreases with an 
increasing value of $m$, and $|\delta S(Q_0,m)|$ becomes 
increasingly smaller as $m$ becomes a large number.  In practice, however, 
$|\delta S(Q_0,m)|$ fluctuates between 0 and a small value $\epsilon$ due to 
the presence of numerical noise at large radial distances, 
which can be reduced by averaging $S(Q)$ [in Eq.\,(\ref{eq6})] 
over many independent sets of fitted Gaussian parameters. 
Further, a value of $R_c$ of about 30 {\AA} is found to be 
sufficient for the calculation of $S(Q)$ from 
Eqs.\,(\ref{eq5}) and (\ref{eq6}). The results from our calculations 
suggest that the average value~\cite{eps} of $\epsilon$ is 
typically of the order of 0.025 for radial distances between 
20 {\AA} and 30 {\AA}. This roughly translates into an error 
of 1.7\%, assuming $S(Q_0)$=1.5 for as-deposited samples from 
experiments~\cite{Xie:2013}.  Thus, aside from small fluctuations 
of $S(Q_0)$ owing to numerical noise, the extended-range 
oscillations in the radial region of 20--30 {\AA} do not 
seem to play any observable role in determining the 
intensity of the FSDP in {\asi}. 
A similar conclusion was reached in a recent 
study~\cite{Dahal:2021}, where an alternative 
argument based on the analysis of the behavior 
of $rG(r)$ and the sampling of $\sin(Qr)/Qr$ 
within the radial region from 0 to $R_c$ in 
Eq.\,(\ref{eq5}) was provided by the authors of 
the study to support this conclusion.

\section{Conclusions}
The present study addresses the origin of the 
extended-range oscillations in {\asi} from a 
real-space point of view. By analyzing a class 
of large partially-ordered networks of Si atoms 
with radial ordering up to a distance of 6 
{\AA} in the PCF, it has been shown that the 
inclusion of short-range ordering in the first 
two coordination shells of the disordered 
networks can lead to an increased ordering of 
the atomic radial distribution in distant 
coordination shells. A comparison of these results with those obtained 
from large {\asi} and disordered crystalline 
configurations reveals that the shell pair-correlation 
functions for the coordination shells of {\asi} at radial 
distances of 20--30 {\AA} are considerably 
ordered and that this 
radial ordering manifests in the form of weak 
oscillations in the total PCF of {\asi}, which 
can be expressed as a sum of the partial radial 
distributions from each coordination shell.  
By using the full width at half maximum of the peak(s) 
of the partial PCFs and the Shannon information
as a measure of the degree of order/disorder, one 
arrives at the conclusion that local atomic 
correlations can considerably affect the distribution 
of atoms in {\asi} up to a distance of 40 {\AA}. 

An analysis of the amplitude of radial 
oscillations in the reduced PCF of 400,000-atom MD 
models of {\asi} shows that the envelope function of 
the reduced PCF decays almost exponentially and 
the resulting decay length (of 4.81 {\AA}) is 
found to be close to the experimental value 
(of 4.23 {\AA}), estimated from the Fourier 
transform of the diffraction data obtained 
for annealed samples of {\asi}.  Likewise, the period of the 
extended-range oscillations (for MD models) is 
found to be about 3.2 {\AA}, which compares well 
with the corresponding experimental value of 3.03 
{\AA} for annealed samples.
The study also shows that the structure factor 
of {\asi} can be expressed as a linear combination 
of a series of Gaussian functions, whose 
amplitude is modulated by a $sinc$ function. 
A convergence study of the intensity of the FSDP, 
using the structure factor obtained from the 
Gaussian approximation, with respect to the number 
of peaks in real space shows that the structure 
of the FSDP is primarily determined by the radial 
correlations originating from a distance of up 
to 20 {\AA} in {\asi} networks, which leads to 
the conclusion that the ERO has no discernible 
effects on the FSDP in {\asi}.

\section{Conflicts of interest} 
There are no conflicts of interest to declare. 

\section{Acknowledgments}
The work was partially supported by the U.S. National 
Science Foundation (NSF) under Grant No.\,DMR 1833035.  
One of us (P.B.) thanks Dr.\, Raymond Atta-Fynn for 
discussions. 

\section{References}
\bibliographystyle{apsrev4-2}

\begin{thebibliography}{54}
\bibitem [{gre()}]{green}%
  \BibitemOpen
  \href@noop {} {}\bibinfo {note} {In the recursion method, the diagonal
  element of $\Ghat(z)$ is computed by transforming the tight-binding $H$
  matrix into a tridiagonal form using the Lanczos algorithm and expressing the
  element as a continued fraction. The coefficients of the continued fraction
  are determined by the diagonal and off-diagonal elements of the tridiagonal
  matrix and a suitable terminator scheme, which preserves the analytic
  properties of $\Ghat(z)$, is used to evaluate the continued
  fraction.}\BibitemShut {Stop}%
\bibitem [{Fis()}]{Fisher}%
  \BibitemOpen
  \href@noop {} {}\bibinfo {note} {The Fisher-Pearson coefficient of skewness
  of a distribution, $p(x)$, is expressed as the standardized third central
  moment, $\mathbf{E[((x-\mu_1)/\nu)^3]}$, where $\mu_n$ is the conventional
  $n$th moment and $\nu^2 = \mu_2 - \mu_1^2$ of $p(x)$. For samples of finite
  size, $n$, one often multiplies by a scale factor of $\sqrt{n(n-1)}/(n-2)$ to
  obtain the adjusted Fisher-Pearson coefficient. The value of the adjusted
  coefficient is reported in this study.}\BibitemShut {Stop}%
\bibitem [{cov()}]{cov}%
  \BibitemOpen
  \href@noop {} {}\bibinfo {note} {The covariance is expressed as the average
  cross product between two variables. One often uses $M$-$k$ instead of $M$ in
  the denominator of Eq.\,(\ref{coreal}). However, this does not affect the
  results for $n \ll M$.}\BibitemShut {Stop}%
\bibitem [{eps()}]{eps}%
  \BibitemOpen
  \href@noop {} {}\bibinfo {note} {Owing to the stochastic nature of the
  minimization problem in fitting Eq.\,(\ref{eq3}) with $G(r)$ of {\asi}, we
  have used here 20 independent sets of Gaussian parameters, $(a_i, b_i, c_i)$,
  to obtain a robust estimate of the average value of $\epsilon$.}\BibitemShut
  {Stop}%
\end{thebibliography}
%

\end{document}